\documentclass[preprint,11pt]{elsarticle}

\usepackage{dcolumn}

\usepackage[utf8]{inputenc}
\usepackage{amsmath}
\usepackage{amsfonts}
\usepackage{graphicx}
\usepackage[export]{adjustbox}
\usepackage[svgnames,dvipsnames]{xcolor}

\usepackage{subcaption}
\usepackage{tikz}
\usepackage{gensymb}
\usetikzlibrary{patterns, decorations.pathreplacing}
\usetikzlibrary{patterns, arrows.meta, calc, decorations.pathreplacing}
\definecolor{waterblue}{RGB}{180, 220, 255} 

\usepackage{natbib} 
\usepackage{amssymb}
\usepackage{amsthm}
\usepackage{graphicx}
\usepackage{epstopdf, epsfig}
\usepackage{bm}

\usepackage{placeins}
\usepackage{pgfplots}
\usepgfplotslibrary{fillbetween}
\pgfplotsset{compat=1.15}
\newlength\figH
\newlength\figW
\usepackage{bm}
\usepackage{overpic} 
\usepackage{tabularx}  

\usepackage{array}     
\usepackage{mathtools}  
\usepackage{adjustbox} 
\pgfplotsset{compat=newest}

\definecolor{cadmiumorange}{rgb}{0.93, 0.53, 0.18}

\usepackage{hyphenat}

\newcommand{\We}{\mathrm{We}}
\renewcommand{\Re}{\mathrm{Re}} 
\newcommand{\Oh}{\mathrm{Oh}}

\newcommand{\De}{\mathrm{De}}
\newcommand{\Wi}{\mathrm{Wi}}

\usepackage{geometry}
\usepackage{booktabs}
\usepackage{caption}
\usepackage{multirow}
\journal{International Journal of Multiphase Flow}

\begin{document}

\begin{frontmatter}



\title{Inelastic spreading of viscoelastic drops} 

\author[label1,label2]{Mete Abbot\corref{cor1}}
\ead{mete.abbot@gmail.com}
\author[label1]{Thijs Varkevisser}
\author[label3]{Manglesh Singh}
\author[label1]{Antoine Gaillard}
\author[label3]{Devranjan Samanta}
\author[label1]{Daniel Bonn}
\ead{d.bonn@uva.nl}

\cortext[cor1]{Corresponding author}

\affiliation[label1]{organization={Van der Waals-Zeeman Institute, Institute of Physics, University of Amsterdam},
             addressline={Science Park 904},
             city={Amsterdam},
             postcode={1098 XH},
             state={},
             country={Netherlands}}

\affiliation[label2]{organization={Department of Aerospace Engineering, Dokuz Eylül University},
             addressline={},
             city={Izmir},
             postcode={35390},
             state={},
             country={Turkey}}

\affiliation[label3]{organization={Department of Mechanical Engineering, Indian Institute of Technology Ropar},
             addressline={},
             city={Rupnagar},
             postcode={140001},
             state={Punjab},
             country={India}}

\begin{abstract}
When a liquid drop impacts a solid surface, rebound and splashing are known to be suppressed by additives that induce elastic effects; however, the influence of elasticity on the maximum spreading radius remains debated. The difficulty lies in isolating elastic resistance from the enhanced spreading caused by viscous shear thinning. Here, we experimentally decouple these effects by investigating Boger fluid drops (constant-viscosity solutions with high elasticity) of polyethylene oxide (PEO) and polyacrylamide (PAAM). We demonstrate that elastic properties do not alter the maximum spreading ratio, even at high concentrations up to 1000 ppm. We formulate an energy balance that accounts for the inertial, capillary, viscous, and elastic contributions, and yields a dimensionless criterion ($\Gamma$) that estimates the degree of elastic effects. We show that $\Gamma \ll 1$ for all impact conditions tested, demonstrating that elastic effects are energetically negligible during droplet spreading, opposite to what Weissenberg and Deborah numbers would predict.
\end{abstract}

\begin{graphicalabstract}
\usetikzlibrary{arrows.meta, positioning, calc}

\definecolor{waterblue}{RGB}{70, 150, 220}
\definecolor{bogerpurple}{RGB}{140, 70, 180}
\definecolor{darkgray}{RGB}{50, 50, 50}
\definecolor{accentorange}{RGB}{235, 120, 35}

\begin{tikzpicture}[font=\sffamily, >=Latex]


  \node[anchor=north west, font=\large\bfseries\color{darkgray}] at (1.5, 7.2) {Identical Spreading};

  \draw[line width=1.5pt, darkgray] (0.5, 4.3) -- (7.5, 4.3);

  \fill[waterblue!45, draw=waterblue!90!black, line width=1pt] 
    (0.8, 4.3) 
    .. controls (0.8, 4.62) and (0.98, 4.58) .. (1.15, 4.45)
    -- (3.05, 4.45)
    .. controls (3.22, 4.58) and (3.4, 4.62) .. (3.4, 4.3) 
    -- cycle;

  \node[font=\bfseries\small\color{waterblue!90!black}] at (2.1, 5.1) {Newtonian Drop};
  \draw[<->, thick, darkgray] (0.8, 3.85) -- (3.4, 3.85) node[midway, below=1pt, font=\scriptsize\bfseries] {$D_{\text{max}(0)}$};

  \fill[bogerpurple!35, draw=bogerpurple!90!black, line width=1pt] 
    (4.5, 4.3) 
    .. controls (4.5, 4.62) and (4.68, 4.58) .. (4.85, 4.45)
    -- (6.75, 4.45)
    .. controls (6.92, 4.58) and (7.1, 4.62) .. (7.1, 4.3) 
    -- cycle;

  \node[font=\bfseries\small\color{bogerpurple!90!black}] at (5.8, 5.1) {Boger Drop};
  \draw[<->, thick, darkgray] (4.5, 3.85) -- (7.1, 3.85) node[midway, below=1pt, font=\scriptsize\bfseries] {$D_{\text{max}(\Gamma)}$};

  \node[draw=red!80!black, fill=red!5, rounded corners=4pt, inner sep=6pt, align=center, line width=0.8pt] at (3.5, 1.9) {
    \small\bfseries Classical Rheology Expectation:\\[2pt]
    $\text{Wi} \gg 1 \text{ and } \text{De} \gg 1$ \\
    $\implies \text{Solid-like Resistance?}$\\
    \small\color{red!80!black}\textbf{Paradox: $D_{\text{max}}$ is Identical!}
  };

  \draw[line width=1pt, darkgray!25] (8.0, 0.4) -- (8.0, 7.2);

  \node[anchor=north west, font=\large\bfseries\color{darkgray}] at (9.6, 7.2) {Energetic Criterion ($\Gamma$)};

  \node[draw=accentorange, fill=accentorange!10, rounded corners=4pt, inner sep=5pt, align=center, line width=1pt] at (12.1, 5.8) {
    $\displaystyle \Gamma \equiv \frac{\text{Elastic Energy}}{\text{Viscous Energy}} = \frac{8}{5\pi} \left( \frac{G D_0}{\mu_i V_0 \sqrt{\mathcal{E}}} \right)$
  };

  \fill[gray!20] (9.0, 2.8) rectangle (13.5, 3.6);
  \draw[darkgray, densely dashed, line width=0.8pt] (9.0, 3.2) -- (15.2, 3.2) node[right, font=\small\bfseries] {1.0};

  \fill[yellow!40, opacity=0.85] (13.5, 1.6) rectangle (15.0, 5.0);
  \node[rotate=90, font=\footnotesize\bfseries, color=darkgray!90!black] at (14.25, 3.3) {Elastic Effects $\Gamma \ge 1$};

  \fill[blue!80!black] (9.4, 3.2) circle (2.8pt);
  \fill[blue!80!black] (9.8, 3.12) circle (2.8pt);
  \fill[red!80!black]  (10.3, 3.28) circle (2.8pt);
  \fill[red!80!black]  (10.8, 3.18) circle (2.8pt);
  \fill[blue!80!black] (11.3, 3.24) circle (2.8pt);

  \node[font=\scriptsize\bfseries, color=blue!80!black, align=center] at (10.6, 2.2) {All Data: $\Gamma \ll 1$};

  \node[fill=darkgray, text=white, rounded corners=3pt, font=\small\bfseries, inner sep=4pt] at (12.1, 0.7) {
    Spreading is Energetically Inelastic ($\Gamma \ll 1$)
  };

  \draw[->, line width=1.1pt] (9.0, 1.6) -- (15.5, 1.6) node[below=2pt, font=\small\bfseries] {$\Gamma$};
  \draw[->, line width=1.1pt] (9.0, 1.6) -- (9.0, 4.5) node[above=-2pt, font=\scriptsize\bfseries, align=center] {$\frac{D_{\text{max}}(\Gamma)}{D_{\text{max}}(0)}$};

\end{tikzpicture}
\end{graphicalabstract}

\begin{highlights}
\item Polymer elasticity does not alter the maximum spreading of drops.
\item Dimensionless parameter $\Gamma$ bounds the observable elastic effects.
\item Elastic storage is energetically negligible during the spreading phase.
\item Weissenberg and Deborah numbers do not predict the spreading.
\item Polymer concentration amplifies thinning faster than elastic resistance.
\end{highlights}


\end{frontmatter}

\section{Introduction}

The impact of viscoelastic drops is critical for applications ranging from agricultural sprays to inkjet printing and forensic blood analysis \cite{josserand2016drop, yarin2006drop,cheng2022drop}. Newtonian spreading is known to be driven by inertia, but resisted by viscous and capillary forces \cite{abbot2026kinematic,laan2014maximum, lee2016universal, de2019predicting,varkevisser2026capillary}. This spreading remains unaffected by the addition of rigid particulates such as nanoparticles \cite{abbot2025nanoparticles}, yet a minimal amount of polymeric macromolecules introduces a high degree of elasticity that can be measured using a conventional rheometer or a Capillary Breakup Rheometer (CaBER) \cite{larson2005rheology,gaillard2024beware,vadillo2010rheological,clasen2006dilute}. Classical scaling dictates that these elastic stresses should resist rapid deformations such as those experienced in drop impact. This expectation became a paradox when Bergeron \textit{et al.} \cite{bergeron2000controlling} showed that trace polymers suppress drop retraction, yet leave the maximum spreading radius identical to that of the Newtonian solvent, as illustrated in Figure \ref{fig:bergeron_paradox}(a) and (b).

\input{Figure1a_code}

The suppression of drop retraction was initially attributed to high elongational viscosity dissipating the impact energy \cite{bergeron2000controlling}. However, this was ruled out because elongational viscosity would equally restrict the initial expansion phase—which it does not—and experiments on small targets without solid boundaries showed no polymer-induced slowdown \cite{rozhkov2003impact, bartolo2007dynamics}. Instead, the suppression of rebound originates from non-Newtonian normal stresses generated specifically at the moving contact line \cite{bartolo2007dynamics}. During retraction, the droplet surface forms a thin liquid wedge with a small receding contact angle, forcing the fluid into extreme local shear rates. In viscoelastic polymer solutions, these high shear gradients generate large first normal stress differences that directly oppose the surface tension driving the retraction \cite{bartolo2007dynamics,kansal2024viscoelastic}. Consequently, the critical threshold for suppressing droplet rebound is dictated by the local Weissenberg number during the retraction phase \cite{dhar2019onset}. Recent work has refined this framework, demonstrating that substituting the dynamic receding contact angle into this elasto-capillary balance predicts the retraction kinematics of viscoelastic drops without relying on phenomenological friction \cite{varkevisser2026}. This extreme sensitivity to the solid boundary is further underscored by findings that modifying substrate flexibility also drastically alters these impact and rebound dynamics \cite{singh2025effect}.

The main question remains open: if polymers cause elasticity that is sufficient to dominate filament thinning \cite{gaillard2022determines}, sprayed sheet breakup \cite{gaillard2022determines,varkevisser2026effect}, splashing \cite{vega2017suppressing},  and drop retraction \cite{bergeron2000controlling,dhar2019onset,singh2025effect}, why does this elasticity not resist droplet spreading? 

Resolving the question experimentally is complicated by several factors. First, high-concentration polymer solutions inherently couple rising elasticity with severe shear thinning \cite{larson2005rheology,gaillard2022determines,varkevisser2026effect}, potentially offsetting any elastic resistance. Recent work successfully modeled this empirically without any elastic considerations \cite{mobaseri2025maximum}. Second, impact shear rates ($\gg 1000~\mathrm{s}^{-1}$) far exceed the limits of standard rheometers, while high-frequency techniques that measure elasticity produce low shear rates \cite{vadillo2010rheological}. Even recently proposed Drop Impact Viscometry (DIV), which promises shear rates relevant to drop impact, cannot explicitly decouple elastic from viscous stresses \cite{abbot2024spreading,abbot2025elucidating,abbot2025minimal,tang2025droplet,piskunov2025spreading}. Finally, dynamic surface tension in aqueous PEO drops decreases from $\sigma \sim 72~\mathrm{mN/m}$ to $\sim 60~\mathrm{mN/m}$ in seconds after the fresh surface is generated \cite{varkevisser2026effect,gaillard2022determines}, adding another coupled variable.

From a rheological perspective, polymer relaxation time $\tau$ relative to the impact timescale ($\tau_{\text{impact}} \approx 10^{-4}$ to $10^{-3}~\mathrm{s}$) produces the Deborah number $\De =\tau/\tau_{\text{impact}} \gg 1$, predicting elastic solid-like behavior. This response was recently suggested to be amplified at $De \approx 1$ \cite{avni2026maximal}. Similarly, the product of the large impact shear rate $\dot\gamma_s$ and relaxation time yields a Weissenberg number $\Wi = 2\dot\gamma_s \tau \gg 1$, which would traditionally suggest elastic dominance, opposite to the observations of Bergeron \textit{et al.}. A recent computational study integrated the viscoelastic energy budgets and numerically estimated the elastic contributions to be less than 1\% for shear-thinning dilute polymer solutions \cite{shende2026forces}. Nevertheless, a generalized physical criterion for the elastic effects based on analytical solutions or scaling arguments and validated by experimental evidence has not yet been established.

In this work, we define the experimental detectability limits of elastic contribution during droplet spreading by providing an energetic boundary condition. We support this framework through experiments on Boger fluids (high elasticity, constant viscosity) of poly(ethylene oxide) (PEO) or polyacrylamide (PAAM) in Newtonian solvents (water, glycerol-water mixtures), enabling the explicit experimental isolation of the elastic contribution from shear-thinning kinematics.

\section{Experimental Details}

The setup consisted of a syringe pump equiped with a blunt needle to produce free-falling drops with constant sizes, 2.72 to 2.83 mm in diameter. The impact target is a hydrophobic Parafilm-coated stainless steel substrate with a static contact angle of $\theta = 110 \pm 5$ degrees. High-speed shadowgraphy imaging was performed using a Phantom TMX 7512 camera, capable of imaging with a temporal resolution of $\sim$ 10 $\mu$s and a spatial resolution of $\sim 10~\mu\text{m/px}$. Further setup details can be found in Supplementary Material section S2.1. 

Aqueous polymer solutions were prepared using polyethylene oxide with molecular weights (PEO 1M, 2M, and 8M) and polyacrylamide (PAAM 5M and 15M). A stock solution of 10,000 ppm (weight parts per million) was produced by mixing for 16 h with a mechanical stirrer until a homogeneous solution was obtained. To prevent aggregation of the polymer upon contact with water, the polymer powder was mixed with an equal mass of isopropanol before adding water. This protocol ensured highly reproducible solutions, as confirmed by shear rheological measurements shown in Figure ~\ref{fig:viscosity_repesentative}. Further shear and extensional rheological characterization details and datasets are provided in the Supplementary Material. 

The solutions are categorized into three distinct rheological regimes:

\begin{enumerate}
    \item \textbf{Boger Fluids:} PAAM in high-viscosity (60, 70, 80, and 90 vol.\%) aqueous glycerol solutions. These fluids possess a constant high-shear viscosity and high elasticity.
    \item \textbf{Quasi-Boger Fluids:} Dilute PEO and PAAM (concentration $\leq 1000$ ppm) in DI water, which have minimal shear-thinning (thinning exponent $n \approx 0.95 - 1.0$), yet possess high elasticity. 
    \item \textbf{Shear Thinning:} Concentrated PEO and PAAM (concentration $2000-10000$ ppm) in purely aqueous solvents. 
\end{enumerate}

As illustrated in the Supplementary Material through dynamic bubble tensiometry measurements, polymer diffusion and adsorption timescales are significantly longer than the impact timescales. Therefore, the surface tension governing spreading kinematics remains close to the pure solvent plateau ($\sigma \approx 72~\text{mN/m}$) for all solutions, minimizing interfacial artifacts.

Further experimental details of the properties of each liquid are tabulated in the Supplementary Materials section S2. In summary, the experiments span a broad range of impact conditions, with Weber numbers ($\We = \rho D_0 V_0^2 / \sigma$) $155 \le \We \le 763$, Reynolds numbers ($\Re = \rho D_0V_0/\mu_0$) $18 \leq \Re \leq 11{,}500$, and Ohnesorge numbers ($\Oh = \mu_0 / \sqrt{\rho \sigma D_0}$) $0.002 \le \Oh \le 1$, where impact parameters are the drop diameter $D_0$ and impact velocity $V_0$, and material properties are density $\rho$, surface tension $\sigma$, and zero shear viscosity $\mu_0$. This range encompasses inertia-dominated, capillary-dominated, and highly viscous impact regimes.

\begin{figure}[t] 
    \centering
    \pgfplotsset{compat=1.17} 

\definecolor{mycolor0}{RGB}{50,155,105}    
\definecolor{mycolor5}{RGB}{213,81,156}    
\definecolor{mycolor10}{RGB}{54,80,163}    
\definecolor{mycolor125}{RGB}{244,196,48}  
\definecolor{mycolor15}{RGB}{245,127,34}   
\definecolor{mycolorSplash}{RGB}{204,204,255} 
\definecolor{mycolorNo}{RGB}{224,224,224}     

\begin{tikzpicture}
\begin{axis}[
width=1.2\figW,
height=1.0\figH,
at={(0\figW,0\figH)},
    xlabel={\small Shear Rate (s$^{-1}$)},
    ylabel={\small $\mu~(\mathrm{mPa\cdot s})$ or $N_1 (\mathrm{Pa})$},
    xmin=1, xmax=10000,
    ymin=0.1, ymax=10000,
    xmode=log,
    ymode=log,
    legend style={at={(1.10,0.90)}, anchor=north west, draw=none, fill=none, font=\footnotesize},
    ticklabel style = {font=\footnotesize}, legend cell align=left, legend image post style={scale=0.8}, 
    row sep=-2pt 
    ]

\addlegendimage{empty legend}
\addlegendentry{\hspace{-0.6cm}\textbf{Viscosity $\mu~(\mathrm{mPa\cdot s})$}}

\addplot[color=teal, line width=2.5pt]
table[row sep=crcr]{
1	242.9	\\
21.4	215.4	\\
41.8	209.9	\\
62.2	206.35	\\
82.6	203.7	\\
103	201.65	\\
123	200.05	\\
144	198.7	\\
164	197.6	\\
184	196.7	\\
205	195.85	\\
225	195.15	\\
246	194.6	\\
266	194.1	\\
286	193.7	\\
307	193.25	\\
327	192.95	\\
348	192.65	\\
368	192.35	\\
388	192.05	\\
409	191.8	\\
429	191.55	\\
450	191.25	\\
470	191.05	\\
490	190.75	\\
511	190.5	\\
531	190.25	\\
551	190.05	\\
572	189.8	\\
592	189.55	\\
613	189.3	\\
633	189.1	\\
653	188.9	\\
674	188.7	\\
694	188.5	\\
715	188.3	\\
735	188.15	\\
755	187.95	\\
776	187.7	\\
796	187.55	\\
817	187.3	\\
837	187.15	\\
857	186.95	\\
878	186.65	\\
898	186.45	\\
918	186.2	\\
939	186	\\
959	185.8	\\
980	185.6	\\
1.00E+03	185.4	\\
};

\addplot[color=red, line width=2.5pt]
table[row sep=crcr]{
1	16	\\
21.4	16.36666667	\\
41.8	15.3	\\
62.2	15.03333333	\\
82.6	14.86666667	\\
103	14.73333333	\\
123	14.6	\\
144	14.56666667	\\
164	14.43333333	\\
184	14.33333333	\\
205	14.3	\\
225	14.23333333	\\
246	14.2	\\
266	14.13333333	\\
286	14.1	\\
307	14.06666667	\\
327	14.03333333	\\
348	14	\\
368	13.96666667	\\
388	13.93333333	\\
409	13.9	\\
429	13.86666667	\\
450	13.86666667	\\
470	13.83333333	\\
490	13.83333333	\\
511	13.8	\\
531	13.8	\\
551	13.8	\\
572	13.76666667	\\
592	13.73333333	\\
613	13.73333333	\\
633	13.73333333	\\
653	13.73333333	\\
674	13.7	\\
694	13.66666667	\\
715	13.66666667	\\
735	13.66666667	\\
755	13.66666667	\\
776	13.63333333	\\
796	13.63333333	\\
817	13.6	\\
837	13.6	\\
857	13.6	\\
878	13.56666667	\\
898	13.56666667	\\
918	13.53333333	\\
939	13.56666667	\\
959	13.53333333	\\
980	13.53333333	\\
1.00E+03	13.53333333	\\
};

\addplot[color=brown, line width=2.5pt]		
table[row sep=crcr]{		
1.032	1.993	\\
1.287	1.933	\\
1.606	1.947	\\
2.002	2.108	\\
2.497	2.1	\\
3.114	2.155	\\
3.884	1.922	\\
4.843	1.918	\\
6.04	2.136	\\
7.532	2.156	\\
9.393	2.11	\\
11.71	1.934	\\
14.61	1.883	\\
18.22	1.826	\\
22.72	1.821	\\
28.33	1.846	\\
35.33	1.812	\\
44.07	1.941	\\
54.95	2.065	\\
68.53	2.042	\\
85.47	1.934	\\
106.6	2.022	\\
132.9	2.014	\\
165.8	1.991	\\
206.7	1.92	\\
257.8	1.872	\\
321.5	1.876	\\
400.9	1.884	\\
500	    1.969	\\
};		

\addplot[color=blue, line width=2.5pt]		
table[row sep=crcr]{		
0.01	208.1	\\
0.01247	204.4	\\
0.01555	192.7	\\
0.0194	191	\\
0.02419	183.1	\\
0.03016	188.6	\\
0.03762	181.3	\\
0.04691	177.3	\\
0.05851	170.6	\\
0.07295	162	\\
0.09098	148.3	\\
0.1135	134.9	\\
0.1415	119.3	\\
0.1765	102.8	\\
0.2201	87.5	\\
0.2744	74.07	\\
0.3423	65.79	\\
0.4268	63.79	\\
0.5323	60.83	\\
0.6638	58.49	\\
0.8278	54.48	\\
1.032	59.18	\\
1.287	59.24	\\
1.606	57.36	\\
2.002	58.47	\\
2.497	57.38	\\
3.114	56.58	\\
3.884	56.44	\\
4.843	56.4	\\
6.04	55.89	\\
7.532	55.36	\\
9.393	54.78	\\
11.71	53.75	\\
14.61	51.91	\\
18.22	51.18	\\
22.72	50.36	\\
28.33	49.75	\\
35.33	50.44	\\
44.07	49.23	\\
54.95	47.91	\\
68.53	46.4	\\
85.47	44.8	\\
106.6	43.1	\\
132.9	41.31	\\
165.8	39.46	\\
206.7	37.55	\\
257.8	35.62	\\
321.5	33.69	\\
400.9	31.79	\\
500	    29.97	\\
};		

\addlegendentry{PAAM 1000 ppm (Gly 90\%)}
\addlegendentry{PAAM 1000 ppm (Gly 60\%)}
\addlegendentry{PEO 1000 ppm (aq)}
\addlegendentry{PEO 10000 ppm (aq)}

\addlegendimage{empty legend}
\addlegendentry{\hspace{-0.6cm}\textbf{Normal Stress $N_1~(\mathrm{Pa})$}}

\addplot[color=black, mark=square*, mark size=3.0pt, line width=0.5pt]
table[row sep=crcr]{
800	0.649361333	\\
1000	2.124191333	\\
1500	12.613184	\\
2000	28.385348	\\
2500	43.053986	\\
3000	49.79561933	\\
4000	91.53015	\\
5000	90.31691933	\\
6000	130.3957607	\\
};

\addplot[color=orange, mark=*, mark size=3.0pt, line width=0.5pt]
table[row sep=crcr]{
200	4.796192	\\
300	8.334957	\\
400	12.321999	\\
500	21.277386	\\
600	32.657723	\\
700	58.49094	\\
};

\addplot[color=violet, mark=triangle*, mark size=3.0pt, line width=0.5pt]
table[row sep=crcr]{
4	11.06938645	\\
6	17.96304526	\\
10	36.18864367	\\
20	53.57154354	\\
30	76.131999	\\
40	103.9937985	\\
60	154.1754349	\\
100	268.4423065	\\
200	560.271179	\\
300	891.8308229	\\
400	1335.895014	\\
};

\addlegendentry{PEO 100 ppm (aq)}
\addlegendentry{PEO 1000 ppm (aq)}
\addlegendentry{PEO 10000 ppm (aq)}

\end{axis}

\end{tikzpicture}
    \caption{\textbf{Representative rheological characterization.} Steady-shear viscosity $\mu$ (horizontal curves) and first normal stress difference $N_1$ (steeply rising curves with markers) as a function of shear rate $\dot{\gamma}$ for selected Boger fluids (PAAM in aqueous glycerol mixtures), quasi-Boger fluids ($\le 1000~\text{ppm}$ aqueous PEO), and shear-thinning fluids ($10{,}000~\text{ppm}$ aqueous PEO).}
    \label{fig:viscosity_repesentative}
\end{figure}

\section{Theoretical Formulation}

To evaluate deviations caused by fluid elasticity, we must establish a purely Newtonian baseline. We follow a similar energy balance presented in \cite{abbot2026kinematic}. We extend the framework to account for the elastic effect. The detailed derivation of each term, as well as the final balance and elastic effects criterion, is presented in the Supplementary Materials. Below, we summarize the steps to obtain the equation for the maximum spreading ratio $\beta_{\text{max}} \equiv D_{\text{max}}/D_0$, where $D_{\text{max}}$ is the maximum diameter of contact between the droplet and the substrate.

The spreading kinematics are governed by equating the initial kinetic and surface energies to the energy partition at the maximum spreading state, namely, surface energy ($\sigma D_{\text{max}}^2$), viscous dissipation (lubrication layer integrating viscous force over time and volume), and elastic energy. An important distinction is that we consider the spreading time $t_s$ and the average spreading velocity $V_s$ as the output of the impact process related by $D_{\mathrm{max}} \equiv t_s V_s$ and not from asymptotic capillary or viscous arguments. 

To establish an upper bound for the elastic energy term, we model the droplet as a linear elastic (Hookean) solid and evaluate the stored strain energy using Hertzian contact mechanics \cite{johnson1987contact}. Integrating the non-linear Hertzian contact force over the geometric deformation of the sphere dictates that the stored energy scales with the fifth power of the maximum spreading ratio $\beta_{\text{max}}$, yielding a scaling of $\sim G\beta_{\text{max}}^5$, where $G$ is the elastic modulus \cite{jana2026impacting,richard2002contact,johnson1987contact}. Normalizing this by the initial kinetic energy introduces the elastic damping parameter, $\Lambda_{el} = \frac{8}{5\pi} (G/\rho V_0^2) \We^{5/2}$. In contrast, standard constitutive models for polymer fluids (e.g., Neo-Hookean, Oldroyd-B, and FENE-P) predict an elastic energy that scales much more weakly, as $\sim G\beta_{\text{max}}^2$ \cite{mousavi2025spreading,figueiredo2014numerical,zhang2024impact,chowdhury2026bridging,jana2026impacting}. Although the Hertz contact assumption may not represent the exact mechanism for a highly spread polymer drop, it is deliberately chosen to ensure a conservative upper-bound estimate that remains consistent up to the purely elastic limit.

Normalizing the energy terms by the initial kinetic energy transforms the initial energy into an energy density term $\mathcal{E}=1 + 12/\We$, and the maximum spreading state energies into $V^2 T^2$, $\Lambda_\nu V^6 T^5$, and $\Lambda_{el} V^5 T^5$ for the normalized surface energy, viscous dissipation, and elastic energy, respectively. This yields the generalized dimensionless energy balance:
\begin{equation}
    \mathcal{E} \approx V^2 T^2 + \Lambda_\nu V^6 T^5 + \Lambda_{el} V^5 T^5
    \label{eq:energy_balance}
\end{equation}
where $V=V_s/V_0$ and $T=t_s/t_{ic}$ are the dimensionless spreading velocity and time normalized by the initial velocity and the inertio-capillary time, respectively ($t_{ic}=\sqrt{\rho D_0^3 / \sigma}$), and $\Lambda_\nu = \We^2 \Oh$ governs the viscous damping \cite{abbot2026kinematic}.

\begin{figure}[t] 
    \centering
    \begin{subfigure}{0.48\textwidth}
        \centering
        \pgfplotsset{compat=1.17} 

\definecolor{mycolor0}{RGB}{50,155,105}    
\definecolor{mycolor5}{RGB}{213,81,156}    
\definecolor{mycolor10}{RGB}{54,80,163}    
\definecolor{mycolor125}{RGB}{244,196,48}  
\definecolor{mycolor15}{RGB}{245,127,34}   
\definecolor{mycolorSplash}{RGB}{204,204,255} 
\definecolor{mycolorNo}{RGB}{224,224,224}     

\begin{tikzpicture}
\begin{axis}[
width=1.2\figW,
height=0.9\figH,
at={(0\figW,0\figH)},
    xlabel={\small $\De = \tau / \tau_{impact}$},
    ylabel={\small $\Wi = 2 \dot{\gamma}_c \tau$},
    xmin=0.01, xmax=1000,
    ymin=0.01, ymax=100000,
    xmode=log,
    ymode=log,
    ytick={0.01, 1, 100,10000},
    axis on top,
    legend style={at={(0.4,0.22)}, anchor=north west, draw=none, fill=none, font=\footnotesize},
    ticklabel style = {font=\footnotesize}, legend cell align=left, legend image post style={scale=0.8}, 
    row sep=-2pt 
    ]

\addplot[only marks, color=teal, mark=triangle*, mark size=3.0pt, line width=0.5pt]
table[row sep=crcr]{
11.03546224	144.8610262	\\
15.13227866	278.6671807	\\
11.9516846	143.4906379	\\
16.44639965	276.9230817	\\
127.6408087	1266.770212	\\
177.408935	2823.387081	\\
7.495373042	125.4479041	\\
10.21685091	247.8667053	\\
4.749690452	104.5431361	\\
6.423735189	217.4903441	\\
4.951271404	100.6137688	\\
6.716959817	202.8952314	\\
37.31551943	771.3519519	\\
51.05714248	1565.738913	\\
3.160304587	83.48757405	\\
4.249997949	171.3018234	\\
1.858388113	59.34383479	\\
2.480085224	119.2679964	\\
1.966808973	59.79878677	\\
2.630507477	122.4713787	\\
14.53913236	441.3540342	\\
19.60251021	919.3940605	\\
};

\addplot[only marks, color=red, mark=o, mark size=2.5pt, line width=0.5pt]
table[row sep=crcr]{
1.135605236	41.34447517	\\
1.506505152	86.33163071	\\
1.293055227	45.12572153	\\
1.718148809	95.1421534	\\
6.251559524	238.4904645	\\
8.352457299	464.5624137	\\
43.99653343	1149.891924	\\
59.33605546	2426.551024	\\
0.798057592	31.72229811	\\
0.949733899	36.16270841	\\
1.257549204	78.69897555	\\
32.95503998	982.6082342	\\
44.18461914	2071.061466	\\
3.549854872	149.1508377	\\
4.718366136	318.679623	\\
0.72571419	28.5718851	\\
0.959478095	58.20126467	\\
0.792275459	31.83039488	\\
1.047776362	63.27145301	\\
23.05084164	763.9026099	\\
30.79177008	1581.909047	\\
0.550010763	22.83824169	\\
0.726487168	48.58787073	\\
0.560690804	23.38974928	\\
0.740653456	48.94437757	\\
14.09973047	503.6486578	\\
18.73484863	1082.03607	\\
2.145057483	99.26294232	\\
2.840921478	216.2944645	\\
0.262681122	11.34931961	\\
0.346478003	24.47262714	\\
0.354050082	16.0087039	\\
0.467324864	32.95948747	\\
8.73690919	363.6598299	\\
11.57901158	751.4762194	\\
0.804937502	39.64375246	\\
1.063619815	86.39886994	\\
0.114739868	5.567115851	\\
0.151070836	11.69198596	\\
0.148738544	6.961780742	\\
0.19589975	15.22015147	\\
4.65847135	211.3378214	\\
6.163981495	449.1263804	\\
0.442159342	21.97249556	\\
0.583829699	49.03777097	\\
2.172177668	104.4215348	\\
2.871478528	229.8845639	\\
};

\addplot[only marks, color=blue, mark=square, mark size=2.5pt, line width=0.5pt]
table[row sep=crcr]{
20.3264305	39.5315343	\\
26.90410727	91.87848325	\\
33.06996424	161.4569644	\\
12.96774857	65.64518454	\\
16.9664973	106.1069207	\\
20.75134762	158.3410637	\\
6.648610857	49.50386468	\\
8.640365773	77.74357212	\\
10.52589271	123.2771073	\\
4.457250592	38.7763019	\\
5.758516696	73.83567	\\
6.990344908	116.4697896	\\
1.889179902	44.44965925	\\
2.378332727	88.05013624	\\
2.842390349	153.8967198	\\
};

\fill[mycolor125!50] 
    (axis cs: 1, 1e-2) -- 
    (axis cs: 1e4, 1e-2) -- 
    (axis cs: 1e4, 1e5) -- 
    (axis cs: 1e-2, 1e5) -- 
    (axis cs: 1e-2, 1) -- 
    (axis cs: 1, 1) -- cycle;

\node at (axis cs: 30, 1) [anchor=center, align=center, rounded corners=2pt, font=\footnotesize, text=black] {Elastic \\ Regime};

\node at (axis cs: 0.1, 0.1) [anchor=center, align=center, rounded corners=2pt, font=\footnotesize, text=black] {Inelastic \\ Regime};

\end{axis}
\end{tikzpicture} 
        \caption{}
        \label{fig:Wi_De}
    \end{subfigure}
    \hfill
    \begin{subfigure}{0.48\textwidth}
        \centering
        \pgfplotsset{compat=1.17} 

\definecolor{mycolor0}{RGB}{50,155,105}    
\definecolor{mycolor5}{RGB}{213,81,156}    
\definecolor{mycolor10}{RGB}{54,80,163}    
\definecolor{mycolor125}{RGB}{244,196,48}  
\definecolor{mycolor15}{RGB}{245,127,34}   
\definecolor{mycolorSplash}{RGB}{204,204,255} 
\definecolor{mycolorNo}{RGB}{224,224,224}     

\begin{tikzpicture}
\begin{axis}[
width=1.2\figW,
height=0.9\figH,
at={(0\figW,0\figH)},
    xlabel={\small $\Lambda_{\nu} \mathcal{E}^{2}$},
    ylabel={\small $\beta_{\text{max}}(\Gamma=0) / \sqrt{\text{We} ~\mathcal{E}}$},
    xmin=30, xmax=2000000,
    ymin=0.01, ymax=1,
    xmode=log,
    ymode=log,
    legend style={at={(0.00,0.5)}, anchor=north west, draw=none, fill=none, font=\footnotesize},
    ticklabel style = {font=\footnotesize}, legend cell align=left, legend image post style={scale=0.8}, 
    row sep=-2pt 
    ]

\addplot[  line width=2.0pt,  domain=0.0001:10E11, samples=200] {(1+x^0.2)^-1};

\addlegendentry{Eq. \cite{abbot2026kinematic}}

\addplot[only marks, color=teal, mark=triangle*, mark size=3.0pt, line width=0.5pt]
table[row sep=crcr]{
12997.1786	0.181850972	\\
77586.49532	0.12906437	\\
21960.75006	0.182095616	\\
131094.4231	0.129446673	\\
96358.3931	0.188596739	\\
575210.2239	0.144076807	\\
6162.455373	0.192675482	\\
36786.70037	0.138588583	\\
2577.026792	0.205505274	\\
15383.52924	0.151676678	\\
3719.882152	0.203636351	\\
22205.79004	0.147974243	\\
8291.303593	0.219133168	\\
49494.83322	0.159758916	\\
1490.193754	0.21544746	\\
8895.692998	0.157919245	\\
739.4946447	0.224158404	\\
4414.404044	0.162612075	\\
934.4523238	0.22291914	\\
5578.201474	0.163245825	\\
1546.216075	0.230831981	\\
9230.117547	0.170481353	\\
};
\addlegendentry{Shear-thinning fluids}

\addplot[only marks, color=red, mark=o, mark size=2.5pt, line width=0.5pt]
table[row sep=crcr]{
430.2514297	0.229456358	\\
2568.380535	0.169667997	\\
504.2008941	0.227487171	\\
3009.820939	0.16902655	\\
656.5816088	0.241768787	\\
3919.455712	0.172667406	\\
1523.807147	0.213787087	\\
9096.347727	0.158917337	\\
318.2067865	0.234064495	\\
351.8201795	0.230900161	\\
2100.186166	0.174301814	\\
851.5392879	0.217917967	\\
5083.253142	0.161891088	\\
416.8060725	0.244965206	\\
2488.118643	0.183228198	\\
302.5205365	0.232007076	\\
1805.892564	0.169442208	\\
307.0023222	0.234585868	\\
1832.646527	0.169242076	\\
605.041073	0.223645052	\\
3611.785127	0.164685233	\\
272.2684828	0.236254263	\\
1625.303307	0.17633392	\\
280.1116079	0.237413315	\\
1672.122744	0.175738656	\\
376.9181795	0.223829538	\\
2250.008364	0.167879642	\\
302.5205365	0.250808597	\\
1805.892564	0.189450238	\\
248.2909292	0.239446671	\\
1482.1696	0.17992346	\\
246.2741256	0.244471907	\\
1470.130317	0.179500105	\\
302.0723579	0.237573738	\\
1803.217167	0.174755959	\\
250.9800006	0.255148026	\\
1498.221979	0.192744713	\\
240.2237149	0.253724276	\\
1434.012465	0.188154514	\\
244.7055006	0.249501153	\\
1460.766429	0.188775436	\\
264.6494471	0.245566791	\\
1579.821569	0.183184579	\\
246.4982149	0.256162699	\\
1471.468015	0.195823661	\\
246.0500363	0.251426224	\\
1468.792618	0.190894778	\\
125.0553845	0.272706374	\\
583.7163698	0.212229946	\\
1791.709687	0.185242103	\\
};
\addlegendentry{Quasi Boger fluids}

\addplot[only marks, color=blue, mark=square, mark size=2.5pt, line width=0.5pt]
table[row sep=crcr]{
18410.88374	0.106930469	\\
88662.26018	0.090941581	\\
273727.0333	0.078993599	\\
5802.421079	0.161392949	\\
27366.08871	0.114697425	\\
84321.53132	0.091876921	\\
2544.787676	0.182320899	\\
11998.51515	0.12773081	\\
36966.33677	0.105475124	\\
1275.592909	0.185793412	\\
6016.904678	0.143303678	\\
18540.43925	0.11801869	\\
125.0553845	0.27094057	\\
583.7163698	0.213989535	\\
1791.709687	0.185777253	\\
};
\addlegendentry{Boger fluids}

\end{axis}
\end{tikzpicture} 
        \caption{}
        \label{fig:spreading_results}
    \end{subfigure}
    \caption{\textbf{Classical rheological prediction versus empirical Newtonian collapse.} (a) Rheological dimensionless numbers based on CaBER relaxation times predict the dominance of elasticity ($\Wi > 1$ and $\De > 1$). (b) Normalized Newtonian maximum spreading ratio $\beta_{\text{max}}(\Gamma=0)/\sqrt{\text{We}~\mathcal{E}}$  evaluated against the viscous damping parameter ($\Lambda_{\nu}\mathcal{E}^2$). Each data point represents the average of at least three independent impact events. Legend applies to both panels.}
\end{figure}

The contribution of elastic resistance to the total flow resistance is dictated by the ratio of elastic and viscous damping terms, $\Gamma$ (see Supplementary Material for full derivation):
\begin{equation}
    \Gamma \equiv \frac{\text{Elastic Energy}}{\text{Viscous Energy}} = \frac{8}{5\pi} \left( \frac{G D_0}{\mu_i V_0 \sqrt{\mathcal{E}}} \right),
    \label{eq:criterion}
\end{equation}
where $\mu_i$ is the viscosity at the equivalent shear rate to the impact, which for a Boger fluid is $\mu_i = \mu_0$. Unlike classical dimensionless parameters, this criterion is unique to drop impact as it couples the available energy density $\mathcal{E}$, the inertial shear rate $D_0/V_0$, and the elastic modulus to viscosity ratio $G/\mu_i$. Furthermore, $\Gamma$ evaluates macroscopic energy dissipation using the bulk apparent viscosity during the impact $\mu_i$, while $\Wi$ and $\De$ evaluations rely on the polymeric contribution ($\mu_p = \mu_0 - \mu_s$, where $\mu_s$ is the viscosity of the Newtonian solvent) to isolate the microscopic relaxation dynamics of the polymer coils.

To solve the energy balance, we assume a viscous lubrication flow within a spreading film and apply a leading-order average velocity closure ($V \approx \sqrt{\mathcal{E}}$, justified in the Supplementary Material). We then use a harmonic mean interpolation across the inertial and viscous asymptotic limits to obtain an approximate analytical expression for the maximum spreading ratio $\beta_{\text{max}}$ from the non-linear energy balance (derivation is provided in the Supplementary Material):
\begin{equation}
    \beta_{\text{max}}(\Gamma) \approx \sqrt{\We\mathcal{E}} \left[ 1 + \left( \Lambda_\nu \mathcal{E}^2 (1 + \Gamma) \right)^{1/5} \right]^{-1}.
    \label{eq:maxspread}
\end{equation}
This formulation isolates fluid elasticity as a direct kinematic perturbation to the viscous baseline. As $\Gamma \to 0$, the equation recovers the purely Newtonian spreading ratio $\beta_{\text{max}}(\Gamma = 0)$ limit \cite{abbot2026kinematic}. Exceeding the value of $\beta_{\text{max}}$ predicted by this baseline indicates a reduction in apparent viscosity (shear-thinning), whereas reduced spreading requires a sufficiently large $\Gamma$.

To properly evaluate both the apparent viscosity $\mu_i$ and classical rheological parameters, the characteristic impact shear rate ($\dot\gamma_s$) must be explicitly defined. By applying volume conservation to the minimum lamella thickness at the maximum spreading state ($h_{\text{min}} \approx 2 D_0^3/3 D_{\text{max}}^2$ \cite{roisman2009inertia}), the average shear rate is $\dot\gamma_s \sim V_s / h_{\text{min}} \approx (3 V_0/2 D_0)V\beta_{\text{max}}^2$. This expression demonstrates that the traditional inertial shear rate estimate ($\dot{\gamma}_i = V_0/D_0$) significantly underestimates the true shear rate for impacts where $\beta_{\text{max}} \gg 1$.

\section{Results and Discussion}

Evaluating our impact experiments using dimensionless scaling from classical rheology would place them within a solid-like elastic regime ($\De > 1$ or $\Wi > 1$), as shown in Figure \ref{fig:Wi_De}. However, these numbers do not directly relate to whether the elastic energy would be sufficient to arrest droplet spreading. Indeed, as shown in Figure \ref{fig:spreading_results}, the impact maximum spreading data collapses onto the purely viscous Newtonian master curve ($\Gamma = 0$) for all Boger and quasi-Boger fluids up to 1000 parts per million (ppm). This empirically shows that neither the polymer's elastic modulus nor a long extensional relaxation time is sufficient to alter the maximum spreading. The ratio of experimental to theoretical spreading yields an average deviation well below 10\%, which is a typical scatter in drop impact experiments (see Supplementary Material). This collapse further validates the Newtonian baseline in \cite{abbot2026kinematic}.

\begin{figure*}[t] 
    \centering
    \begin{subfigure}{0.43\textwidth}
        \centering
        \pgfplotsset{compat=1.17} 
			
\definecolor{mycolor0}{RGB}{50,155,105}    
\definecolor{mycolor5}{RGB}{213,81,156}    
\definecolor{mycolor10}{RGB}{54,80,163}    
\definecolor{mycolor125}{RGB}{244,196,48}  
\definecolor{mycolor15}{RGB}{245,127,34}   
\definecolor{mycolorSplash}{RGB}{204,204,255} 
\definecolor{mycolorNo}{RGB}{224,224,224}     

\begin{tikzpicture}
\begin{axis}[
    width=1.4\figW,
    height=0.9\figH,
    at={(0\figW,0\figH)},
    xmode=log,
    xmin=0.01, xmax=1000,
    ymin=0, ymax=1.2,
    xlabel={ $\Gamma$},
    ylabel={$\beta_{\text{max}}(\Gamma) / \beta_{\text{max}}(0)$},
    legend style={at={(0.02,0.65)}, anchor=north west, draw=none, fill=none, font=\footnotesize},
    ticklabel style = {font=\footnotesize}, legend cell align=left, legend image post style={scale=0.8}, 
    row sep=-2pt,
    axis on top
    ]

\fill[gray!50] (axis cs:0.01, 0.9) rectangle (axis cs:1000, 1.1);

\fill[mycolor125!50] (axis cs:1, 0) rectangle (axis cs:1000, 0.9);

\addplot[black, dashed, thick, forget plot] coordinates {(0.01, 1) (1000, 1)};

\addplot[thick, color=blue!50!black, domain=0.01:1000, samples=200] 
    {(1 + 1^0.2)/(1 + 1^0.2 * (1 + x)^0.2)};
\addlegendentry{$\Lambda_\nu \mathcal{E}^2 = 1$}

\addplot[thick, color=green!50!black, domain=0.01:1000, samples=200] 
    {(1 + 100^0.2)/(1 + 100^0.2 * (1 + x)^0.2)};
\addlegendentry{$\Lambda_\nu \mathcal{E}^2 = 10^{2}$}

\addplot[thick, color=red!50!black, domain=0.01:1000, samples=200] 
    {(1 + 10000^0.2)/(1 + 10000^0.2 * (1 + x)^0.2)};
\addlegendentry{$\Lambda_\nu \mathcal{E}^2 = 10^{4}$}

\addplot[thick, color=black, dashed, domain=0.01:1000, samples=200] 
    {(1 + x)^(-0.2)};
\addlegendentry{Asymptote}

\addplot[area legend, fill=gray!20, draw=none] coordinates {(0.01,-1)};
\addlegendentry{$\pm 10\%$}

\node at (axis cs: 15, 0.25) [anchor=center, text=black, font=\footnotesize, align=center] {Measurable \\ Elastic Regime};

\end{axis}
\end{tikzpicture} 
        \caption{}
        \label{fig:Gamma_sensitivity}
    \end{subfigure}
    \hfill
    \begin{subfigure}{0.25\textwidth}
        \centering
        \pgfplotsset{compat=1.17} 

\definecolor{mycolor0}{RGB}{50,155,105}    
\definecolor{mycolor5}{RGB}{213,81,156}    
\definecolor{mycolor10}{RGB}{54,80,163}    
\definecolor{mycolor125}{RGB}{244,196,48}  
\definecolor{mycolor15}{RGB}{245,127,34}   
\definecolor{mycolorSplash}{RGB}{204,204,255} 
\definecolor{mycolorNo}{RGB}{224,224,224}     

\begin{tikzpicture}
\begin{axis}[
width=0.9\figW,
height=0.9\figH,
at={(0\figW,0\figH)},
    xlabel={\small $\dot\gamma_s$},
    ylabel={\small $\Gamma$},
    xmin=1000, xmax=30000,
    ymin=0.00001, ymax=100,
    xmode=log,
    ymode=log,
    xtick={1000, 10000}, 
    ytick={0.0001, 0.01, 1, 100},
    axis on top,
    legend style={at={(0.02,0.30)}, anchor=north west, draw=none, fill=none, font=\footnotesize},
    ticklabel style = {font=\footnotesize}, legend cell align=left, legend image post style={scale=0.8}, 
    row sep=-2pt 
    ]

\addplot[only marks, color=teal, mark=triangle*, mark size=3.0pt, line width=0.5pt]
table[row sep=crcr]{
3935.459518	0.205994125	\\
7658.001695	0.131786394	\\
3946.055399	0.060957445	\\
7703.436556	0.038998014	\\
4232.84686	0.00185235	\\
9543.13087	0.001185056	\\
4417.912456	0.325845253	\\
8829.937561	0.208462114	\\
5025.857161	0.519463447	\\
10576.46022	0.332330906	\\
4934.859725	0.179934628	\\
10066.4184	0.115114621	\\
5714.52889	0.010763657	\\
11733.6448	0.006886136	\\
5523.914604	0.628824173	\\
11464.96789	0.402295307	\\
5979.629129	0.724100562	\\
12156.49309	0.463249141	\\
5913.694903	0.286514611	\\
12251.43305	0.18330002	\\
6340.976678	0.023087264	\\
13361.53924	0.014770262	\\
};

\addplot[only marks, color=red, mark=o, mark size=2.5pt, line width=0.5pt]
table[row sep=crcr]{
6265.624872	0.622273921	\\
13234.34931	0.398104731	\\
6158.543582	0.26550354	\\
13134.4708	0.169858018	\\
6956.081504	0.027184663	\\
13706.40166	0.017391606	\\
5439.101182	0.017570087	\\
11610.34926	0.011240604	\\
6519.815117	0.420692228	\\
6344.72303	0.190249352	\\
13967.10957	0.121713548	\\
5651.324906	0.019650755	\\
12048.93352	0.012571728	\\
7141.229737	0.021411576	\\
15434.31448	0.013698227	\\
6405.700953	0.309754129	\\
13199.14892	0.198167688	\\
6548.892939	0.152616086	\\
13167.98777	0.097637365	\\
5952.272222	0.019359633	\\
12468.43781	0.012385481	\\
6642.376743	0.196669289	\\
14294.68038	0.125820754	\\
6707.711018	0.095581274	\\
14198.33209	0.061148887	\\
5962.09639	0.017758114	\\
12956.83096	0.011360896	\\
7485.986425	0.011800157	\\
16500.34551	0.007549245	\\
6823.101082	0.10783086	\\
14882.58304	0.068985657	\\
7112.497359	0.054356958	\\
14812.62876	0.034775299	\\
6716.779014	0.011079061	\\
14039.98723	0.007087918	\\
7747.268803	0.007111702	\\
17079.20587	0.004549768	\\
7661.049057	0.055726023	\\
16275.41278	0.03565117	\\
7408.142247	0.0273527	\\
16383.0097	0.017499109	\\
7176.347608	0.006322851	\\
15426.96692	0.004045095	\\
7809.009993	0.003620503	\\
17629.21834	0.002316246	\\
7522.900936	0.002720323	\\
16752.93191	0.001740349	\\
};

\addplot[only marks, color=blue, mark=square, mark size=2.5pt, line width=0.5pt]
table[row sep=crcr]{
682.7150677	0.001601706	\\
1612.939949	0.001068727	\\
2851.052258	0.000806254	\\
1412.359778	0.004217169	\\
2323.681313	0.002861674	\\
3489.66478	0.002159925	\\
1761.838444	0.009737531	\\
2816.52584	0.00660815	\\
4494.699819	0.004987813	\\
1840.423906	0.020175465	\\
3566.922706	0.013690153	\\
5662.298809	0.010332885	\\
3294.491799	0.164387959	\\
6659.880198	0.111839478	\\
11725.68012	0.084494549	\\
};

\fill[mycolor125!50] (0,1) rectangle (30000,100);

\node at (axis cs: 5500, 10) [anchor=center, text=black, font=\footnotesize, align=center] {Elastic Regime};

\node at (axis cs: 5500, 0.0001) [anchor=center, text=black, font=\footnotesize, align=center] {Zimm model};

\end{axis}
\end{tikzpicture} 
        \caption{}
        \label{fig:Gamma_zimm}
    \end{subfigure}
    \hfill
    \begin{subfigure}{0.25\textwidth}
        \centering
        \pgfplotsset{compat=1.17} 

\definecolor{mycolor0}{RGB}{50,155,105}    
\definecolor{mycolor5}{RGB}{213,81,156}    
\definecolor{mycolor10}{RGB}{54,80,163}    
\definecolor{mycolor125}{RGB}{244,196,48}  
\definecolor{mycolor15}{RGB}{245,127,34}   
\definecolor{mycolorSplash}{RGB}{204,204,255} 
\definecolor{mycolorNo}{RGB}{224,224,224}     

\begin{tikzpicture}
\begin{axis}[
width=0.9\figW,
height=0.9\figH,
at={(0\figW,0\figH)},
    xlabel={\small $\dot\gamma_s$},
    ylabel={},
    xmin=1000, xmax=30000,
    ymin=0.00001, ymax=100,
    xmode=log,
    ymode=log,
    xtick={1000, 10000}, 
    ytick=\empty,
    axis on top,
    legend style={at={(0.02,0.30)}, anchor=north west, draw=none, fill=none, font=\footnotesize},
    ticklabel style = {font=\footnotesize}, legend cell align=left, legend image post style={scale=0.8}, 
    row sep=-2pt 
    ]

\addplot[only marks, color=teal, mark=triangle*, mark size=3.0pt, line width=0.5pt]
table[row sep=crcr]{
4090.195803	0.026685128	\\
7778.442089	0.017072025	\\
4101.208298	0.027205944	\\
7824.591522	0.017405221	\\
4399.275964	0.003331978	\\
9693.219429	0.002131659	\\
4591.618058	0.033919185	\\
8968.809449	0.021700071	\\
5223.466223	0.043871715	\\
10742.80035	0.028067281	\\
5128.890906	0.046069388	\\
10224.73691	0.02947326	\\
5939.215478	0.007204555	\\
11918.18443	0.004609172	\\
5741.106528	0.056186302	\\
11645.28193	0.035945637	\\
6214.739055	0.070192558	\\
12347.68303	0.044906252	\\
6146.212396	0.075139497	\\
12444.11614	0.048071096	\\
6590.294241	0.012278647	\\
13571.68141	0.007855363	\\
};

\addplot[only marks, color=red, mark=o, mark size=2.5pt, line width=0.5pt]
table[row sep=crcr]{
6511.979716	0.072579177	\\
13442.4911	0.04643311	\\
6400.688152	0.075780923	\\
13341.04177	0.048481453	\\
7229.584053	0.019202681	\\
13921.96761	0.012285069	\\
5652.958371	0.004032438	\\
11792.94977	0.002579785	\\
6776.164335	0.06075893	\\
6594.187894	0.063665691	\\
14186.77578	0.040730636	\\
5873.52641	0.004235677	\\
12238.43181	0.002709808	\\
7422.012034	0.02212643	\\
15677.05599	0.014155561	\\
6657.563376	0.058095243	\\
13406.73711	0.037166898	\\
6806.385453	0.055537401	\\
13375.08588	0.035530498	\\
6186.306517	0.004903526	\\
12664.53382	0.00313707	\\
6903.544898	0.051440071	\\
14519.49842	0.032909198	\\
6971.448017	0.057326748	\\
14421.63483	0.036675247	\\
6196.516956	0.004797425	\\
13160.60812	0.003069191	\\
7780.32403	0.019542256	\\
16759.85291	0.012502315	\\
7091.375043	0.058569903	\\
15116.64726	0.037470565	\\
7392.149942	0.040002026	\\
15045.59277	0.025591617	\\
6980.872553	0.004765776	\\
14260.79962	0.003048943	\\
8051.879634	0.020927417	\\
17347.81723	0.013388483	\\
7962.269858	0.092379207	\\
16531.38257	0.059100338	\\
7699.419137	0.089605093	\\
16640.67171	0.057325577	\\
7458.510686	0.005201558	\\
15669.59287	0.003327738	\\
8116.048394	0.032292558	\\
17906.47996	0.020659423	\\
7818.689963	0.006426841	\\
17016.41184	0.004111623	\\
};

\addplot[only marks, color=blue, mark=square, mark size=2.5pt, line width=0.5pt]
table[row sep=crcr]{
682.7150677	0.009822391	\\
1612.939949	0.007022057	\\
2851.052258	0.005297484	\\
1412.359778	0.018564581	\\
2323.681313	0.012597498	\\
3489.66478	0.009508299	\\
1761.838444	0.037339601	\\
2816.52584	0.025339654	\\
4494.699819	0.0191263	\\
1840.423906	0.046994101	\\
3566.922706	0.031888059	\\
5662.298809	0.024068075	\\
3294.491799	0.103764317	\\
6659.880198	0.070594873	\\
11725.68012	0.053334315	\\
};

\fill[mycolor125!50] (0,1) rectangle (30000,100);

\node at (axis cs: 5500, 10) [anchor=center, text=black, font=\footnotesize, align=center] {Elastic Regime};

\node at (axis cs: 5500, 0.0001) [anchor=center, text=black, font=\footnotesize, align=center] {CaBER};

\end{axis}
\end{tikzpicture} 
        \caption{}
        \label{fig:Gamma_caber}
    \end{subfigure}
    \caption{\textbf{Evaluation of the elastic criterion $\Gamma$.} (a) Sensitivity of the relative spreading ratio $\beta_{max}(\Gamma) / \beta_{max}(0)$ evaluated against the energetic criterion $\Gamma$. Shaded region shows typical scatter of drop impact experiments of around 10 \%. (b, c) Evaluation of the experimental $\Gamma$ criterion against the characteristic impact shear rate $\dot\gamma_s \approx (3 V_0/2 D_0)V\beta_{\text{max}}^2$, where $\Gamma$ is calculated using (b) theoretical Zimm entropic relaxation times and (c) CaBER extensional relaxation times. Legend: \textcolor{blue}{$\square$} Boger fluids, {\scriptsize\textcolor{red}{$\bigcirc$}} quasi-Boger fluids, and \textcolor{teal}{$\blacktriangle$} shear-thinning fluids.}
    \label{fig:Gamma_comparison}
\end{figure*}

Given that the general spreading ratio (Eq.~\ref{eq:maxspread}) explicitly depends on elasticity, we aim to determine when this effect becomes experimentally measurable. Figure \ref{fig:Gamma_comparison}(a) maps the theoretical sensitivity of the relative spreading ratio. The horizontal baseline at unity represents purely viscous, Newtonian spreading, surrounded by a gray band indicating the typical $\pm 10\%$ scatter inherent to drop impact experiments \cite{laan2014maximum,lee2016universal,de2019predicting,abbot2026kinematic,abbot2024spreading}. For an elastic resistance to be experimentally observable in a drop impact experiment, the deviation below the Newtonian baseline must exceed the noise floor, which happens when $\Gamma \approx 1$ as shown in Figure \ref{fig:Gamma_comparison}(a). Therefore, the experimentally detectable 'Elastic Regime' (yellow region) only emerges when $\Gamma \ge 1$. Note that if the elastic effect is negligible under Hertzian scaling ($\sim \beta_{\mathrm{max}}^5$), it is mathematically guaranteed to be negligible for other models for polymer elasticity (e.g., Neo-Hookean, Oldroyd-B, and FENE-P) which predict a weaker elastic effect that scales as $\sim G\beta_{\mathrm{max}}^2$ \cite{mousavi2025spreading,figueiredo2014numerical,zhang2024impact,chowdhury2026bridging,jana2026impacting}.

For inertia-dominated impacts ($\We \gg 1$, $\mathcal{E} \approx 1$), solving $\Gamma = 1$ via Equation \ref{eq:criterion} yields a direct material threshold for this boundary: 
\begin{equation}
    G_{\text{critical}} \approx 2\mu_i \left(\frac{V_0}{D_0}\right).
\end{equation}

To observe macroscopic elastic resistance, the elastic modulus of the fluid must be approximately twice the magnitude of its viscous stress. However, because the impact of the drop imposes extreme deformation rates, the characteristic viscous stress dominates the available structural elasticity ($G < G_{\text{critical}}$). As a result, $\Gamma < 1$ holds throughout this work's experimental space shown in Figure \ref{fig:Gamma_comparison}(b) and (c), regardless of whether the elastic modulus $G=\mu_p/\tau$ is evaluated by Zimm relaxation times (representing the theoretical lower bound for time which corresponds to the upper bound for $G$) \cite{clasen2006dilute} or CaBER measurements (representing the experimentally achievable upper bound for time which corresponds to the lower bound for $G$) \cite{campo2010slow, gaillard2022determines}. In all our experiments, which represent a wide range of impact parameters and polymer solutions, the elastic energy storage is energetically negligible and kinematically suppressed below the measurement noise floor. Recent simulations on aqueous Xanthan gum (up to 3000 ppm) found the elastic energy storage to be below 1\% of the impact energy \cite{shende2026forces}, consistent with the energetic hierarchy proposed here and suggesting that the framework proposed is applicable to viscoelastic systems beyond PEO and PAAM.

Evaluating the appropriate elastic modulus $G$ and relaxation time $\tau$ for dilute solutions remains challenging because the exact degree of polymer stretching during drop impact is unknown. The theoretical Zimm framework ($G_Z = \mu_p/\tau= n k_B T$, where $n$ is the polymer chain number density, $k_B$ is the Boltzmann constant, and $T$ is absolute temperature) dictates the lower bound for the relaxation time of isolated coils \cite{clasen2006dilute}, whereas CaBER measurements capture the upper bound of an interacting, highly stretched polymer network \cite{campo2010slow, gaillard2022determines}.


\begin{figure}[t] 
    \centering
    \begin{tikzpicture}

\begin{axis}[
    width=0.9\figW,
    height=0.8\figH,
    scale only axis, 
    xmode=log, ymode=log,
    xmin=0.01, xmax=100,
    ymin=0.1, ymax=100,
    xlabel={$c/c^*$},
    ylabel={$G/\mu_0$ (s$^{-1}$)},
    axis y line*=left,
    axis x line*=bottom,
    y axis line style={black, thick},
    y tick label style={font=\footnotesize},
    x tick label style={font=\footnotesize},
    xlabel style={font=\footnotesize},
    ylabel style={font=\footnotesize},
    legend style={at={(0.03,0.05)}, anchor=south west, font=\footnotesize, draw=none, fill=none}, 
    legend cell align=left,
]

\addplot[only marks, mark=*, red, fill=white, mark size=2.5pt] coordinates { 
    (0.034, 22.445)
    (0.084, 9.490)
    (0.17, 17.512)
    (0.34, 13.813)
    (0.51, 15.165)
    (0.84, 13.953)
    (1.7, 13.213)
    (3.4, 11.927)
};
\addlegendentry{Quasi Boger fluids}

\addplot[only marks, mark=triangle*, teal, fill=white, mark size=3pt] coordinates { 
    (8.4, 7.159)
    (34, 1.969)
};
\addlegendentry{Shear-thinning fluids}

\addplot[smooth, dashed, thick, black] coordinates {
    (0.01, 16.5)
    (0.1, 16.2)
    (0.5, 15.5)
    (1, 14.5)
    (3, 11)
    (10, 5)
    (34, 1.8)
    (100, 0.8)
};
\addlegendentry{$G/\mu_0 = 15.8 /( 1 + \frac{c/c^*}{7.3}^{1.16}$)}

\addlegendimage{thick, black} 
\addlegendentry{$G$}

\end{axis}

\begin{axis}[
    width=0.9\figW,
    height=0.8\figH,
    scale only axis, 
    xmode=log, ymode=log,
    xmin=0.01, xmax=100,
    ymin=0.001, ymax=100000,
    axis y line*=right,
    axis x line*=top, 
    xtick=\empty,     
    x axis line style={black, thick}, 
    ylabel={{$G$ (Pa)}},
    y axis line style={ thick},
    y tick label style={ font=\footnotesize},
    ylabel style={font=\footnotesize},
]

\addplot[thick, black, mark=none] coordinates {
    (0.034, 23.276)
    (0.084, 10.420)
    (0.17, 20.489)
    (0.34, 17.681)
    (0.51, 21.838)
    (0.84, 24.000)
    (1.7, 38.980)
    (3.4, 75.143)
    (8.4, 272.059)
    (34, 29527.560)
};

\addplot[only marks, mark=*, red, mark size=2.5pt] coordinates {
    (0.034, 23.276)
    (0.084, 10.420)
    (0.17, 20.489)
    (0.34, 17.681)
    (0.51, 21.838)
    (0.84, 24.000)
    (1.7, 38.980)
    (3.4, 75.143)
};

\addplot[only marks, mark=triangle*, teal, mark size=3pt] coordinates {
    (8.4, 272.059)
    (34, 29527.560)
};

\end{axis}
\end{tikzpicture}
    \caption{Increasing elastic modulus $G$ (solid symbols, right axis) and its decreasing elastic modulus to zero-shear viscosity $G/\mu_0$ (open symbols, left axis) as a function of PEO-4M polymer concentration in water. $c$ is the polymer mass
    concentration, and $c^*$ = 0.295 kg/m$^3$ (equivalently 296 ppm) is the critical overlap concentration. $G$ data are derived from CaBER relaxation time measurements from \cite{gaillard2022determines}.}
    \label{fig:modulus_ratio}
\end{figure}

As shown in Figure \ref{fig:spreading_results}, concentrated shear-thinning fluids spread beyond the baseline. Increasing the polymer concentration to 10,000 ppm only amplifies this deviation. Counterintuitively, this massive addition of polymer does not increase the relative elastic resistance. As shown in Figure \ref{fig:modulus_ratio}, while the absolute elastic modulus $G$ increases in the concentrated regime, $\mu_0$ and $\mu_i$ increase even faster. As a result, $G/\mu_0$ (or $G/\mu_i$ for thinning) and $\Gamma$ decrease. Thus, the relative elastic effect remains negligible for highly concentrated drops, leaving apparent viscosity reductions (shear thinning) as the primary driver of increased spreading. Therefore, a regime where the spreading is resisted by elasticity is inaccessible using higher concentrations of polymers such as PEO or PAAM.

It must be noted that the thinning behavior stems from the structural breakdown of the polymer network, while the $\Gamma$ framework is based on $\mu_0$. If the impact shear rate exceeds the critical thinning threshold ($\dot{\gamma}_{rheo}^*$), the polymer chains align and disentangle. This reduces the effective viscosity, permitting the droplet to spread beyond the zero-shear prediction \cite{mobaseri2025maximum}. Evaluating this yield boundary requires a characteristic impact shear rate $\dot\gamma_s$ introduced earlier in this work. We define the boundary of yielding using the impact Carreau number $Cu_i = \dot{\gamma}_s/\dot{\gamma}_{rheo}^*$. For Boger fluids and dilute solutions ($Cu_i < 1$), the spreading adheres strictly to the Newtonian baseline governed by the zero-shear viscosity. In contrast, for concentrated solutions ($Cu_i \gg 1$), the reduced apparent viscosity dominates, driving the enhanced spreading observed in Figure \ref{fig:spreading_results} through shear-thinning kinematics.

\section{Conclusion}

In this work, we investigated the apparent contradiction regarding the influence of fluid elasticity on the maximum spreading of impacting droplets. By deriving a dimensionless elasto-viscocapillary energy balance, we provide a physical explanation for why fluid elasticity does not resist droplet spreading. This framework establishes an energetic criterion, Eq.~\ref{eq:criterion}, which evaluates the ratio of elastic to viscous damping penalties for a given impact energy budget. Because impact events impose extreme deformation rates, viscous dissipation dominates the structural elastic capacity of the polymer network ($\Gamma < 1$). Ultimately, the impact-driven spreading remains governed primarily by inertia, capillarity, and viscous dissipation, while the elasticity is energetically negligible for dilute and semi-dilute polymer solutions.

\textit{Data Availability} — The datasets generated and analyzed during the current study are available on arXiv through the link [link will be updated after the review].

\textit{Acknowledgments} — 
This work is supported by the European Research Council (ERC) under the European Union’s Horizon Europe research and innovation programme (Grant agreement No. 101142159).

\textit{CRediT Authorship Contribution Statement} — 
\textbf{Mete Abbot:} Conceptualization, Methodology, Formal analysis, Investigation, Data curation, Visualization, Writing – original draft. 
\textbf{Thijs Varkevisser:} Investigation, Data curation. 
\textbf{Manglesh Singh:} Investigation, Data curation. 
\textbf{Antoine Gaillard:} Conceptualization, Writing – review \& editing. 
\textbf{Devranjan Samanta:} Resources, Supervision, Funding acquisition, Writing – review \& editing. 
\textbf{Daniel Bonn:} Conceptualization, Resources, Supervision, Funding acquisition, Writing – review \& editing.

\textit{Conflict of Interest} — The authors report no conflicts of interest.

\textit{Generative AI statement} — During the preparation of this work, the author(s) used Google Gemini AI and Grammarly for checking grammer and language. The author(s) reviewed and edited the output as needed and take full responsibility for the content of the published article. 

\bibliographystyle{elsarticle-num} 
\bibliography{bibliography}

\begin{thebibliography}{10}
\expandafter\ifx\csname url\endcsname\relax
  \def\url#1{\texttt{#1}}\fi
\expandafter\ifx\csname urlprefix\endcsname\relax\def\urlprefix{URL }\fi
\expandafter\ifx\csname href\endcsname\relax
  \def\href#1#2{#2} \def\path#1{#1}\fi

\bibitem{josserand2016drop}
C.~Josserand, S.~T. Thoroddsen, Drop impact on a solid surface, Annual review of fluid mechanics 48~(1) (2016) 365--391.

\bibitem{yarin2006drop}
A.~L. Yarin, Drop impact dynamics: splashing, spreading, receding, bouncing…, Annu. Rev. Fluid Mech. 38~(1) (2006) 159--192.

\bibitem{cheng2022drop}
X.~Cheng, T.-P. Sun, L.~Gordillo, Drop impact dynamics: Impact force and stress distributions, Annual Review of Fluid Mechanics 54~(1) (2022) 57--81.

\bibitem{abbot2026kinematic}
M.~Abbot, D.~Bonn, Kinematic closure of drop impact, arXiv preprint arXiv:2605.11797 (2026).

\bibitem{laan2014maximum}
N.~Laan, K.~G. De~Bruin, D.~Bartolo, C.~Josserand, D.~Bonn, Maximum diameter of impacting liquid droplets, Physical Review Applied 2~(4) (2014) 044018.

\bibitem{lee2016universal}
J.~B. Lee, N.~Laan, K.~G. de~Bruin, G.~Skantzaris, N.~Shahidzadeh, D.~Derome, J.~Carmeliet, D.~Bonn, Universal rescaling of drop impact on smooth and rough surfaces, Journal of Fluid Mechanics 786 (2016) R4.

\bibitem{de2019predicting}
T.~C. de~Goede, K.~G. de~Bruin, N.~Shahidzadeh, D.~Bonn, Predicting the maximum spreading of a liquid drop impacting on a solid surface: effect of surface tension and entrapped air layer, Physical review fluids 4~(5) (2019) 053602.

\bibitem{varkevisser2026capillary}
T.~Varkevisser, D.~Bonn, Capillary-viscous retraction dynamics of droplets: The role of the dynamic contact angle, Physical Review Fluids 11~(1) (2026) 013601.

\bibitem{abbot2025nanoparticles}
M.~Abbot, M.~H. Iqbal, L.~Liu, E.~Koos, I.~V. Roisman, J.~Hussong, A.~A. Castrej{\'o}n-Pita, J.~R. Castrej{\'o}n-Pita, Nanoparticles do not influence droplet break-up, spreading, or splashing, Journal of Colloid and Interface Science 693 (2025) 137570.

\bibitem{larson2005rheology}
R.~G. Larson, The rheology of dilute solutions of flexible polymers: Progress and problems, Journal of Rheology 49~(1) (2005) 1--70.

\bibitem{gaillard2024beware}
A.~Gaillard, M.~A. Herrada, A.~Deblais, J.~Eggers, D.~Bonn, Beware of caber: Filament thinning rheometry does not always give ‘the’relaxation time of polymer solutions, Physical Review Fluids 9~(7) (2024) 073302.

\bibitem{vadillo2010rheological}
D.~Vadillo, T.~Tuladhar, A.~Mulji, M.~Mackley, The rheological characterization of linear viscoelasticity for ink jet fluids using piezo axial vibrator and torsion resonator rheometers, Journal of Rheology 54~(4) (2010) 781--795.

\bibitem{clasen2006dilute}
C.~Clasen, J.~Plog, W.-M. Kulicke, M.~Owens, C.~Macosko, L.~Scriven, M.~Verani, G.~H. McKinley, How dilute are dilute solutions in extensional flows?, Journal of Rheology 50~(6) (2006) 849--881.

\bibitem{bergeron2000controlling}
V.~Bergeron, D.~Bonn, J.~Y. Martin, L.~Vovelle, Controlling droplet deposition with polymer additives, Nature 405~(6788) (2000) 772--775.

\bibitem{rozhkov2003impact}
A.~Rozhkov, B.~Prunet-Foch, M.~Vignes-Adler, Impact of drops of polymer solutions on small targets, Physics of Fluids 15~(7) (2003) 2006--2019.

\bibitem{bartolo2007dynamics}
D.~Bartolo, A.~Boudaoud, G.~Narcy, D.~Bonn, Dynamics of non-newtonian droplets, Physical review letters 99~(17) (2007) 174502.

\bibitem{kansal2024viscoelastic}
M.~Kansal, V.~Bertin, C.~Datt, J.~Eggers, J.~H. Snoeijer, Viscoelastic wetting: Cox--voinov theory with normal stress effects, Journal of fluid mechanics 985 (2024) A17.

\bibitem{dhar2019onset}
P.~Dhar, S.~R. Mishra, D.~Samanta, Onset of rebound suppression in non-newtonian droplets post-impact on superhydrophobic surfaces, Physical Review Fluids 4~(10) (2019) 103303.

\bibitem{varkevisser2026}
T.~Varkevisser, On droplets and sprays, Ph.d. thesis, Universiteit van Amsterdam (2026).

\bibitem{singh2025effect}
M.~Singh, S.~Basu, D.~Samanta, Effect of substrate flexibility on non-newtonian droplet impact dynamics on flexible superhydrophobic surfaces, Journal of Colloid and Interface Science 679 (2025) 100--112.
\newblock \href {https://doi.org/10.1016/j.jcis.2025.02.012} {\path{doi:10.1016/j.jcis.2025.02.012}}.

\bibitem{gaillard2022determines}
A.~Gaillard, R.~Sijs, D.~Bonn, What determines the drop size in sprays of polymer solutions?, Journal of Non-Newtonian Fluid Mechanics 305 (2022) 104813.

\bibitem{varkevisser2026effect}
T.~Varkevisser, D.~Bonn, Effect of high wind speeds on droplet formation in sprays of dilute polymer solutions, Journal of Non-Newtonian Fluid Mechanics (2026) 105618.

\bibitem{vega2017suppressing}
E.~Vega, A.~Castrej{\'o}n-Pita, Suppressing prompt splash with polymer additives, Experiments in Fluids 58~(5) (2017) 57.

\bibitem{mobaseri2025maximum}
A.~Mobaseri, S.~Kumar, X.~Cheng, Maximum spreading of impacting shear-thinning and shear-thickening drops, Proceedings of the National Academy of Sciences 122~(22) (2025) e2500163122.

\bibitem{abbot2024spreading}
M.~Abbot, M.~Lannert, A.~Kiran, S.~Bakshi, J.~Hussong, I.~V. Roisman, Spreading of a viscous drop after impact onto a spherical target, Journal of Fluid Mechanics 996 (2024) A10.

\bibitem{abbot2025elucidating}
M.~Abbot, F.~Yin, M.~Piskunov, I.~V. Roisman, J.~Hussong, Elucidating on the drop impact dynamics of water-in-oil microemulsions via advanced rheometry, Colloids and Surfaces A: Physicochemical and Engineering Aspects 720 (2025) 136952.

\bibitem{abbot2025minimal}
M.~Abbot, F.~Yin, A.~L. Azevedo~Costa, I.~V. Roisman, J.~Hussong, Minimal addition of fumed nanoparticles suppresses droplet splashing, Applied Physics Letters 127~(8) (2025).

\bibitem{tang2025droplet}
S.~Tang, X.~Li, W.~Wang, W.~Zhao, Y.~Zhou, S.~Wang, Y.~Li, P.~Yu, X.~Wen, G.~Hu, et~al., Droplet impact-based microliter viscometry, Analytical Chemistry 97~(25) (2025) 13076--13085.

\bibitem{piskunov2025spreading}
M.~Piskunov, X.~Zhang, W.~Yuan, F.~Chen, L.~Chen, M.~Abbot, Spreading of water-in-oil emulsion drop on oleophilic surface, Physics of Fluids 37~(5) (2025).

\bibitem{avni2026maximal}
O.~Avni, D.~Wang, M.~Ravisankar, R.~Zenit, Maximal spreading of impacting viscoelastic droplets, arXiv preprint arXiv:2601.15246 (2026).

\bibitem{shende2026forces}
T.~Shende, E.~Moeendarbary, I.~Eames, Forces and energy of viscoelastic droplets impacting a wall, Journal of Rheology 70~(5) (2026) 995--1021.
\newblock \href {https://doi.org/10.1122/8.0001202} {\path{doi:10.1122/8.0001202}}.

\bibitem{johnson1987contact}
K.~L. Johnson, Contact mechanics, Cambridge university press, 1987.

\bibitem{jana2026impacting}
S.~Jana, J.~Kolinski, D.~Lohse, V.~Sanjay, Impacting spheres: from liquid drops to elastic beads, Soft matter 22~(11) (2026) 2226--2236.

\bibitem{richard2002contact}
D.~Richard, C.~Clanet, D.~Qu{\'e}r{\'e}, Contact time of a bouncing drop, Nature 417~(6891) (2002) 811--811.

\bibitem{mousavi2025spreading}
M.~Mousavi, S.~A. Faroughi, Spreading and rebound of viscoelastic droplets on surfaces with hybrid wettability, Physics of Fluids 37~(1) (2025).

\bibitem{figueiredo2014numerical}
R.~A. Figueiredo, C.~M. Oishi, J.~A. Cuminato, J.~C. Azevedo, A.~M. Afonso, M.~A. Alves, Numerical investigation of three dimensional viscoelastic free surface flows: impacting drop problem, in: 11th World Congress on Computational Mechanics, WCCM, 2014.

\bibitem{zhang2024impact}
H.~Zhang, Q.~Luan, W.~Yuan, F.~Chen, B.~Meng, Impact and spread dynamics of a viscoelastic droplet on an inclined hydrophilic surface, Physics of Fluids 36~(8) (2024).

\bibitem{chowdhury2026bridging}
A.~Chowdhury, S.~Mitra, S.~K. Mitra, Bridging liquid and elastic solid impact regimes using flexible hydrogels, Langmuir (2026).

\bibitem{roisman2009inertia}
I.~V. Roisman, Inertia dominated drop collisions. ii. an analytical solution of the navier--stokes equations for a spreading viscous film, Physics of Fluids 21~(5) (2009).

\bibitem{campo2010slow}
L.~Campo-Deano, C.~Clasen, The slow retraction method (srm) for the determination of ultra-short relaxation times in capillary breakup extensional rheometry experiments, Journal of Non-Newtonian Fluid Mechanics 165~(23-24) (2010) 1688--1699.

\end{thebibliography}

\end{document}


\title{Supplementary Material for the article: \\ Inelastic spreading of viscoelastic drops}

\author[label1,label2]{Mete Abbot\corref{cor1}}
\ead{mete.abbot@gmail.com}
\author[label1]{Thijs Varkevisser}
\author[label3]{Manglesh Singh}
\author[label1]{Antoine Gaillard}
\author[label3]{Devranjan Samanta}
\author[label1]{Daniel Bonn}
\ead{d.bonn@uva.nl}

\cortext[cor1]{Corresponding author}

\affiliation[label1]{organization={Van der Waals-Zeeman Institute, Institute of Physics, University of Amsterdam},
             addressline={Science Park 904},
             city={Amsterdam},
             postcode={1098 XH},
             state={},
             country={Netherlands}}

\affiliation[label2]{organization={Department of Aerospace Engineering, Dokuz Eylül University},
             addressline={},
             city={Izmir},
             postcode={35390},
             state={},
             country={Turkey}}

\affiliation[label3]{organization={Department of Mechanical Engineering, Indian Institute of Technology Ropar},
             addressline={},
             city={Rupnagar},
             postcode={140001},
             state={Punjab},
             country={India}}

\date{\today}

\maketitle

\setcounter{section}{0}
\renewcommand{\thesection}{S\arabic{section}}
\renewcommand{\theequation}{S\arabic{equation}}
\renewcommand{\thefigure}{S\arabic{figure}}
\renewcommand{\thetable}{S\arabic{table}}

\tableofcontents
\newpage

\section{Nomenclature}

\begin{table}[h]
    \centering
    \caption{Nomenclature of the Roman Symbols used in the main manuscript and in the supplementary.}
    \vspace{0.2cm}
    \renewcommand{\arraystretch}{1.1} 
    \begin{tabular}{@{}ll@{}}
        \toprule
        \multicolumn{2}{@{}l}{\textbf{Roman Symbols}} \\
        \midrule
        $c$ & Polymer mass concentration \\
        $c^*$ & Critical overlap concentration \\
        $D_0$ & Initial droplet diameter \\
        $D_{\text{max}}$ & Maximum spreading diameter \\
        $E_{k}$ & Initial kinetic energy \\
        $E_{s,0}$ & Initial surface energy \\
        $E_{s,max}$ & Final surface energy at maximum spreading \\
        $E_{el}$ & Stored elastic strain energy \\
        $E_{total}$ & Total initial energy \\
        $G$ & Elastic modulus \\
        $G_{\text{critical}}$ & Critical elastic modulus threshold \\
        $G_Z$ & Theoretical entropic (Zimm) elastic modulus \\
        $h$ & Instantaneous film thickness \\
        $h_{\text{min}}$ & Minimum film thickness at maximum spread \\
        $k_B$ & Boltzmann constant \\
        $K$ & Hertzian stiffness for an incompressible solid \\
        $M_w$ & Molecular weight \\
        $n$ & Polymer chain density / thinning exponent \\
        $N_1$ & First normal stress difference \\
        $N_A$ & Avogadro's constant \\
        $R_0$ & Initial droplet radius \\
        $t_{ic}$ & Inertio-capillary time ($t_{ic} = \sqrt{\rho D_0^3 / \sigma}$) \\
        $t_s$ & Spreading time \\
        $T$ & Absolute temperature / Dimensionless time ($t_s/t_{ic}$) \\
        $V_0$ & Initial impact velocity \\
        $V_s$ & Spreading velocity \\
        $V$ & Dimensionless spreading velocity ($V_s/V_0$) \\
        $W_v$ & Total viscous work \\
        $\dot{W}$ & Instantaneous viscous dissipation rate \\
        \bottomrule
    \end{tabular}
    \label{tab:nomenclature_roman}
\end{table}

\begin{table}[h]
    \centering
    \caption{Nomenclature of the Greek Symbols used in the main manuscript and in the supplementary.}
    \vspace{0.2cm}
    \renewcommand{\arraystretch}{1.1} 
    \begin{tabular}{@{}ll@{}}
        \toprule 
        \multicolumn{2}{@{}l}{\textbf{Greek Symbols}} \\
        \midrule
        $\beta_{\text{max}}$ & Maximum spreading ratio ($D_{\text{max}}/D_0$) \\
        $\Gamma$ & Elastic effects criterion (ratio of elastic to viscous damping) \\
        $\dot{\gamma}$ & Shear rate \\
        $\dot\gamma_s$ & Characteristic impact shear rate \\
        $\dot{\gamma}_{c, rheo}$ & Critical thinning threshold shear rate \\
        $\delta_{\text{max}}$ & Geometric maximum deformation \\
        $\mathcal{E}$ & Total available energy density \\
        $\theta$ & Static/receding contact angle \\
        $\Lambda_{\nu}$ & Viscous damping parameter \\
        $\Lambda_{el}$ & Elastic damping parameter \\
        $\Lambda_{eff}$ & Effective damping parameter \\
        $\mu$ & Apparent shear viscosity \\
        $\mu_0$ & Zero-shear viscosity \\
        $\mu_i$ & Apparent viscosity at equivalent impact shear rate \\
        $\mu_p$ & Polymeric contribution to viscosity ($\mu_0 - \mu_s$) \\
        $\mu_s$ & Solvent viscosity \\
        $\nu$ & Poisson's ratio \\
        $\El$ & Elasto-inertial number ($G/\rho V_0^2$) \\
        $\rho$ & Fluid density \\
        $\sigma$ & Surface tension \\
        $\tau$ & Extensional relaxation time \\
        $\tau_{\text{impact}}$ & Impact timescale \\
        $\tau_Z$ & Theoretical Zimm relaxation time \\
        $\Omega$ & Droplet volume \\   
        \bottomrule
    \end{tabular}
    \label{tab:nomenclature_greek}
\end{table}

\clearpage

\section{Experimental method}

\subsection{Experimental Setup}

A schematic of the experimental setup used in this study is shown in Figure~\ref{f1}. It consists of a syringe pump, a vertical stage, a stainless steel spherical target, and a high-speed Phantom TMX7510 high-speed camera (temporal resolution $\sim$ 10 $\mu$s, spatial resolution $\sim 10~\mu\text{m/px}$). All mounted on a passive pneumatic vibration isolation table. Harvard Apparatus syringe pump (Darwin Microfluidics) is set to a flow rate of 0.3 ml/min with a 21G blunt bevel facet needle, with an external diameter of 0.8 mm and an internal diameter of 0.51 mm, allowing the generation of drops with a consistent size, 2.72 to 2.83 mm in diameter. The size of the drop is determined by the balance of surface tension and the weight of the drop. Impact velocities are determined by changing the vertical impact distance between the needle tip and the substrate. The height and position are adjusted using a coarse vertical stage, followed by finer adjustment with an XYZ-microstage. Velocities up to 5.0 m/s are achieved. A protective acrylic tube helps to stabilize the trajectory of the drop. 

The camera has an image-based auto trigger that engages when the drop arrives in the proximity of the target. In order to achieve shadowgraphy high-speed imaging, backlighting LED illumination is placed behind the drop. A diffuser sheet with a diffusion angle of 80\textdegree ~is placed between the LED illumination and the impact zone of the drop to achieve more uniform illumination of the falling drop. 

The impact target is a hydrophobic Parafilm-coated stainless steel substrate with a static contact angle of $\theta = 110 \pm 5$ degrees.

To ensure statistical reliability, every experimental condition was repeated a minimum of three times. The kinematic parameters ($D_0$, $V_0$, $D_{max}$) extracted from these high-speed image sequences were ensemble-averaged to determine the representative spreading behavior. The relative standard deviation among the triplicate measurements for individual conditions was typically below 5 \%.

\begin{figure}[h]
    \centering

\begin{tikzpicture}[
    font=\sffamily\small,
    >=stealth,
    label line/.style={draw, gray!80, thick, -},
]

\filldraw[fill=gray!15, draw=gray!70, thick] (-6.5, -1.5) rectangle (6.5, -0.5);
\node[gray!80!black] at (0, -1.0) {\textbf{Pneumatic Vibration Isolation Table}};

\filldraw[fill=gray!40, draw=gray!80!black, thick] (-0.8, -0.5) rectangle (0.8, 0);

\filldraw[fill=gray!30, draw=gray!80!black, thick] (-0.1, 4.5) rectangle (0.1, 5.5); 
\filldraw[fill=gray!10, draw=gray!80!black, thick, rounded corners=2pt] (-1.2, 5.5) rectangle (1.2, 6.5);
\node[align=center] at (0, 6) {Syringe Pump \&\\Z-stage};

\draw[dashed, draw=blue!50, thick, fill=blue!5, fill opacity=0.3] (-0.4, 1.4) rectangle (0.4, 4.5);

\filldraw[fill=cyan!50, draw=cyan!80!black, thick] (0, 3.2) circle (0.15);
\draw[->, thick, cyan!80!black] (0.7, 3.2) -- (0.7, 2.2) node[midway, right] {$V_0$};

\filldraw[fill=gray!80!black, draw=black, thick, rounded corners=2pt] (3.5, 0.2) rectangle (5.5, 1.8);
\filldraw[fill=gray!60, draw=black, thick] (3.0, 0.6) rectangle (3.5, 1.4); 
\filldraw[fill=cyan!20, draw=black, thick] (2.9, 0.7) rectangle (3.0, 1.3); 
\node[text=white, align=center] at (4.5, 1) {High-Speed\\Camera};

\filldraw[fill=yellow!10, draw=orange!80!black, thick, rounded corners=2pt] (-5.5, 0.2) rectangle (-4.0, 1.8);
\node[align=center, orange!80!black] at (-4.75, 1) {LED\\Backlight};
\filldraw[fill=gray!10, draw=gray!50, thick] (-3.3, -0.2) rectangle (-3.1, 2.2); 

\draw[->, yellow!80!orange, thick, dashed] (-3.9, 1) -- (-3.4, 1);
\draw[->, yellow!80!orange, thick, dashed] (-3.9, 1.5) -- (-3.4, 1.5);
\draw[->, yellow!80!orange, thick, dashed] (-3.9, 0.5) -- (-3.4, 0.5);

\draw[label line] (-3.2, 2.2) -- (-3.2, 2.8) node[above, text=black] {Diffuser ($80^\circ$)};

\draw[label line] (-0.4, 2.5) -- (-1.3, 2.5) node[left, text=black, align=right] {Acrylic\\Tube};

\draw[label line] (-0.1, 5) -- (-1.5, 5) node[left, text=black, align=right] {Blunt Needle\\(ID: $0.51$ mm)};

\draw[label line] (-0.15, 3.2) -- (-1.5, 3.7) node[left, text=black, align=right] {Droplet};

\draw[label line] (0, 0) -- (0, 0) node[above, text=black, align=left] {Parafilm Coating\\($\theta = 110^\circ \pm 5^\circ$)};

\end{tikzpicture}
    \caption{Experimental setup.}
    \label{f1}
\end{figure}  

\subsection{Rheological and Interfacial Characterization}

\subsubsection{Shear Rheology}

To characterize the flow behavior of the polymer solutions, steady shear rheometry was performed. Figure \ref{fig:viscosity_boger} presents the apparent shear viscosity, $\mu$, as a function of the shear rate, $\dot{\gamma}$, for the Boger fluids. Figure \ref{fig:viscosity_boger} presents the same parameters for a representative PEO ($M_w = 1 \times 10^6$ g/mol) solution at various concentrations. While the highly concentrated solutions exhibit slight shear thinning, the dilute solutions of 1000 ppm or below maintain a near-constant viscosity plateau across the measured shear rates, effectively approximating Newtonian viscous behavior. 

However, the non-Newtonian character of these dilute solutions is revealed through their elastic response. As shown in Figure \ref{fig:n1_peo_4m}, the first normal stress difference, $N_1$, increases significantly with the applied shear rate. The measurable $N_1$ confirms that these fluids store elastic strain energy under deformation. This validates their classification as highly elastic fluids, ensuring that any absence of elastic resistance during drop impact cannot be attributed to a lack of intrinsic fluid elasticity.

\begin{figure}[h] 
    \centering
    \pgfplotsset{compat=1.17} 

\definecolor{mycolor0}{RGB}{50,155,105}    
\definecolor{mycolor5}{RGB}{213,81,156}    
\definecolor{mycolor10}{RGB}{54,80,163}    
\definecolor{mycolor125}{RGB}{244,196,48}  
\definecolor{mycolor15}{RGB}{245,127,34}   
\definecolor{mycolorSplash}{RGB}{204,204,255} 
\definecolor{mycolorNo}{RGB}{224,224,224}     

\begin{tikzpicture}
\begin{axis}[
width=1.2\figW,
height=1.0\figH,
at={(0\figW,0\figH)},
    xlabel={\small Shear Rate (s$^{-1}$)},
    ylabel={\small Viscosity (mPa$\cdot$s)},
    xmin=0, xmax=1000,
    ymin=0.1, ymax=1000,
    ymode=log,
    legend style={at={(1.10,0.90)}, anchor=north west, draw=none, fill=none, font=\footnotesize},
    ticklabel style = {font=\footnotesize}, legend cell align=left, legend image post style={scale=0.8}, 
    row sep=-2pt 
    ]

\addplot[color=teal, mark=triangle*, mark size=2.0pt, line width=0.5pt]
table[row sep=crcr]{
1	242.9	\\
21.4	215.4	\\
41.8	209.9	\\
62.2	206.35	\\
82.6	203.7	\\
103	201.65	\\
123	200.05	\\
144	198.7	\\
164	197.6	\\
184	196.7	\\
205	195.85	\\
225	195.15	\\
246	194.6	\\
266	194.1	\\
286	193.7	\\
307	193.25	\\
327	192.95	\\
348	192.65	\\
368	192.35	\\
388	192.05	\\
409	191.8	\\
429	191.55	\\
450	191.25	\\
470	191.05	\\
490	190.75	\\
511	190.5	\\
531	190.25	\\
551	190.05	\\
572	189.8	\\
592	189.55	\\
613	189.3	\\
633	189.1	\\
653	188.9	\\
674	188.7	\\
694	188.5	\\
715	188.3	\\
735	188.15	\\
755	187.95	\\
776	187.7	\\
796	187.55	\\
817	187.3	\\
837	187.15	\\
857	186.95	\\
878	186.65	\\
898	186.45	\\
918	186.2	\\
939	186	\\
959	185.8	\\
980	185.6	\\
1.00E+03	185.4	\\
};
\addlegendentry{90\% Glycerol}

\addplot[color=red, mark=square*, mark size=2.0pt, line width=0.5pt]
table[row sep=crcr]{
1	100.2	\\
21.4	84.45	\\
41.8	81.9	\\
62.2	80.1	\\
82.6	78.8	\\
103	77.8	\\
123	77	\\
144	76.35	\\
164	75.8	\\
184	75.3	\\
205	74.95	\\
225	74.6	\\
246	74.3	\\
266	74.1	\\
286	73.9	\\
307	73.7	\\
327	73.5	\\
348	73.4	\\
368	73.3	\\
388	73.1	\\
409	73	\\
429	72.9	\\
450	72.8	\\
470	72.7	\\
490	72.55	\\
511	72.5	\\
531	72.4	\\
551	72.35	\\
572	72.25	\\
592	72.15	\\
613	72.05	\\
633	72	\\
653	71.9	\\
674	71.85	\\
694	71.8	\\
715	71.8	\\
735	71.75	\\
755	71.75	\\
776	71.75	\\
796	71.7	\\
817	71.75	\\
837	71.7	\\
857	71.7	\\
878	71.7	\\
898	71.7	\\
918	71.7	\\
939	71.6	\\
959	71.6	\\
980	71.6	\\
1.00E+03	71.5	\\
};
\addlegendentry{80\% Glycerol}

\addplot[color=orange, mark=diamond*, mark size=2.0pt, line width=0.5pt]
table[row sep=crcr]{
1	44.56666667	\\
21.4	37.4	\\
41.8	36.33333333	\\
62.2	35.56666667	\\
82.6	35.06666667	\\
103	34.56666667	\\
123	34.2	\\
144	33.9	\\
164	33.6	\\
184	33.4	\\
205	33.2	\\
225	33	\\
246	32.8	\\
266	32.7	\\
286	32.6	\\
307	32.5	\\
327	32.4	\\
348	32.3	\\
368	32.2	\\
388	32.13333333	\\
409	32.1	\\
429	32.03333333	\\
450	32	\\
470	31.96666667	\\
490	31.93333333	\\
511	31.9	\\
531	31.83333333	\\
551	31.83333333	\\
572	31.8	\\
592	31.76666667	\\
613	31.76666667	\\
633	31.73333333	\\
653	31.73333333	\\
674	31.7	\\
694	31.7	\\
715	31.63333333	\\
735	31.63333333	\\
755	31.6	\\
776	31.6	\\
796	31.6	\\
817	31.53333333	\\
837	31.53333333	\\
857	31.56666667	\\
878	31.53333333	\\
898	31.53333333	\\
918	31.5	\\
939	31.53333333	\\
959	31.5	\\
980	31.53333333	\\
1.00E+03	31.46666667	\\
};
\addlegendentry{70\% Glycerol}

\addplot[color=blue, mark=*, mark size=2.0pt, line width=0.5pt]
table[row sep=crcr]{
1	16	\\
21.4	16.36666667	\\
41.8	15.3	\\
62.2	15.03333333	\\
82.6	14.86666667	\\
103	14.73333333	\\
123	14.6	\\
144	14.56666667	\\
164	14.43333333	\\
184	14.33333333	\\
205	14.3	\\
225	14.23333333	\\
246	14.2	\\
266	14.13333333	\\
286	14.1	\\
307	14.06666667	\\
327	14.03333333	\\
348	14	\\
368	13.96666667	\\
388	13.93333333	\\
409	13.9	\\
429	13.86666667	\\
450	13.86666667	\\
470	13.83333333	\\
490	13.83333333	\\
511	13.8	\\
531	13.8	\\
551	13.8	\\
572	13.76666667	\\
592	13.73333333	\\
613	13.73333333	\\
633	13.73333333	\\
653	13.73333333	\\
674	13.7	\\
694	13.66666667	\\
715	13.66666667	\\
735	13.66666667	\\
755	13.66666667	\\
776	13.63333333	\\
796	13.63333333	\\
817	13.6	\\
837	13.6	\\
857	13.6	\\
878	13.56666667	\\
898	13.56666667	\\
918	13.53333333	\\
939	13.56666667	\\
959	13.53333333	\\
980	13.53333333	\\
1.00E+03	13.53333333	\\
};
\addlegendentry{60\% Glycerol}

\addplot[color=brown, mark=*, mark size=2.0pt, line width=0.5pt]
table[row sep=crcr]{
1	1.9	\\
21.4	1.9	\\
41.8	1.9	\\
62.2	1.9	\\
82.6	1.9	\\
103	1.9	\\
123	1.9	\\
144	1.9	\\
164	1.9	\\
184	1.9	\\
205	1.9	\\
225	1.9	\\
246	1.9	\\
266	1.9	\\
286	1.9	\\
307	1.9	\\
327	1.9	\\
348	1.9	\\
368	1.9	\\
388	1.9	\\
409	1.9	\\
429	1.9	\\
450	1.8	\\
470	1.8	\\
490	1.8	\\
511	1.8	\\
531	1.8	\\
551	1.8	\\
572	1.8	\\
592	1.8	\\
613	1.8	\\
633	1.8	\\
653	1.8	\\
674	1.8	\\
694	1.8	\\
715	1.8	\\
735	1.8	\\
755	1.9	\\
776	1.9	\\
796	1.9	\\
817	1.9	\\
837	1.9	\\
857	1.9	\\
878	1.9	\\
898	1.9	\\
918	1.9	\\
939	2	\\
959	1.9	\\
980	2	\\
1.00E+03	2	\\
};
\addlegendentry{Water}

\end{axis}
\end{tikzpicture}
    \caption{Viscosity $\mu$ as a function of shear rate $\dot{\gamma}$ of the Boger fluids tested. The legend indicates the solvent of the PAAM at 1000 ppm in glycerol-water solutions.}
    \label{fig:viscosity_boger}
\end{figure}

\begin{figure}[h] 
    \centering
    \pgfplotsset{compat=1.17} 

\definecolor{mycolor0}{RGB}{50,155,105}    
\definecolor{mycolor5}{RGB}{213,81,156}    
\definecolor{mycolor10}{RGB}{54,80,163}    
\definecolor{mycolor125}{RGB}{244,196,48}  
\definecolor{mycolor15}{RGB}{245,127,34}   
\definecolor{mycolorSplash}{RGB}{204,204,255} 
\definecolor{mycolorNo}{RGB}{224,224,224}     

\begin{tikzpicture}
\begin{axis}[
    width=1.2\figW, 
    height=1.0\figH,
    xlabel={\small Shear Rate (s$^{-1}$)},
    ylabel={\small Viscosity (Pa$\cdot$s)},
    xmin=1, xmax=1000,
    ymin=0.0001, ymax=0.5,
    xmode=log,
    ymode=log,
    legend style={at={(1.05,0.95)}, anchor=north west, draw=none, fill=none, font=\footnotesize},
    ticklabel style = {font=\footnotesize}, 
    legend cell align=left, 
    legend image post style={scale=0.8}, 
    row sep=-2pt 
    ]

\addplot[color=teal, mark=triangle*, mark size=2.0pt, line width=0.5pt]		
table[row sep=crcr]{		
0.01	0.2081	\\
0.01247	0.2044	\\
0.01555	0.1927	\\
0.0194	0.191	\\
0.02419	0.1831	\\
0.03016	0.1886	\\
0.03762	0.1813	\\
0.04691	0.1773	\\
0.05851	0.1706	\\
0.07295	0.162	\\
0.09098	0.1483	\\
0.1135	0.1349	\\
0.1415	0.1193	\\
0.1765	0.1028	\\
0.2201	0.0875	\\
0.2744	0.07407	\\
0.3423	0.06579	\\
0.4268	0.06379	\\
0.5323	0.06083	\\
0.6638	0.05849	\\
0.8278	0.05448	\\
1.032	0.05918	\\
1.287	0.05924	\\
1.606	0.05736	\\
2.002	0.05847	\\
2.497	0.05738	\\
3.114	0.05658	\\
3.884	0.05644	\\
4.843	0.0564	\\
6.04	0.05589	\\
7.532	0.05536	\\
9.393	0.05478	\\
11.71	0.05375	\\
14.61	0.05191	\\
18.22	0.05118	\\
22.72	0.05036	\\
28.33	0.04975	\\
35.33	0.05044	\\
44.07	0.04923	\\
54.95	0.04791	\\
68.53	0.0464	\\
85.47	0.0448	\\
106.6	0.0431	\\
132.9	0.04131	\\
165.8	0.03946	\\
206.7	0.03755	\\
257.8	0.03562	\\
321.5	0.03369	\\
400.9	0.03179	\\
500	0.02997	\\
};		
\addlegendentry{PEO 1M 10000 ppm}		
		
\addplot[color=black, mark=x, mark size=2.0pt, line width=0.5pt]		
table[row sep=crcr]{		
0.01	0.04761	\\
0.01247	0.07179	\\
0.01556	0.07178	\\
0.01939	0.07237	\\
0.02419	0.08156	\\
0.03016	0.08675	\\
0.03762	0.0865	\\
0.04691	0.08903	\\
0.0585	0.08685	\\
0.07296	0.08087	\\
0.09098	0.07654	\\
0.1135	0.07008	\\
0.1415	0.06121	\\
0.1765	0.05211	\\
0.2201	0.04183	\\
0.2745	0.03394	\\
0.3423	0.03044	\\
0.4268	0.02909	\\
0.5323	0.02967	\\
0.6638	0.02865	\\
0.8278	0.02822	\\
1.032	0.02881	\\
1.287	0.02838	\\
1.606	0.02692	\\
2.002	0.02814	\\
2.497	0.02795	\\
3.114	0.02738	\\
3.884	0.02744	\\
4.843	0.02745	\\
6.04	0.02746	\\
7.532	0.02731	\\
9.393	0.02702	\\
11.71	0.02694	\\
14.61	0.02625	\\
18.22	0.026	\\
22.72	0.02533	\\
28.33	0.02509	\\
35.33	0.02479	\\
44.07	0.02458	\\
54.95	0.02442	\\
68.53	0.02461	\\
85.47	0.02417	\\
106.6	0.02355	\\
132.9	0.02287	\\
165.8	0.02214	\\
206.7	0.02137	\\
257.8	0.02055	\\
321.5	0.01972	\\
400.9	0.01886	\\
500	0.018	\\
};		
\addlegendentry{PEO 1M 7500 ppm}			
		
\addplot[color=pink, mark=diamond, mark size=2.0pt, line width=0.5pt]		
table[row sep=crcr]{		
0.01247	0.05723	\\
0.01555	0.1291	\\
0.01939	0.1442	\\
0.02419	0.1382	\\
0.03016	0.1263	\\
0.03762	0.09692	\\
0.04691	0.07464	\\
0.0585	0.07819	\\
0.07296	0.0716	\\
0.09098	0.06495	\\
0.1135	0.05828	\\
0.1415	0.04935	\\
0.1765	0.0388	\\
0.2201	0.03019	\\
0.2744	0.02576	\\
0.3423	0.02416	\\
0.4268	0.02248	\\
0.5323	0.0195	\\
0.6638	0.01173	\\
0.8278	0.01234	\\
1.032	0.01448	\\
1.287	0.01338	\\
1.606	0.01363	\\
2.002	0.01251	\\
2.497	0.01264	\\
3.114	0.01157	\\
3.884	0.01264	\\
4.843	0.01282	\\
6.04	0.01225	\\
7.532	0.01233	\\
9.393	0.01222	\\
11.71	0.01228	\\
14.61	0.01174	\\
18.22	0.01177	\\
22.72	0.01146	\\
28.33	0.01144	\\
35.33	0.01148	\\
44.07	0.01152	\\
54.95	0.0117	\\
68.53	0.01135	\\
85.47	0.01091	\\
106.6	0.0109	\\
132.9	0.01079	\\
165.8	0.01091	\\
206.7	0.01067	\\
257.8	0.0104	\\
321.5	0.01012	\\
400.9	0.009827	\\
500	0.009514	\\
};		
\addlegendentry{PEO 1M 5000 ppm}

\addplot[color=magenta, mark=oplus, mark size=2.0pt, line width=0.5pt]		
table[row sep=crcr]{		
0.009994	0.09219	\\
0.01247	0.09425	\\
0.01555	0.08349	\\
0.0194	0.07505	\\
0.02419	0.06857	\\
0.03016	0.06692	\\
0.03761	0.05802	\\
0.04691	0.05068	\\
0.0585	0.04616	\\
0.07296	0.03924	\\
0.09098	0.03257	\\
0.1135	0.02912	\\
0.1415	0.02622	\\
0.1765	0.02305	\\
0.2201	0.0217	\\
0.2744	0.01837	\\
0.3423	0.01365	\\
0.4268	0.009728	\\
0.5323	0.006651	\\
0.6638	0.004089	\\
0.8278	0.003914	\\
1.032	0.004491	\\
1.287	0.005022	\\
1.606	0.00371	\\
2.002	0.004438	\\
2.497	0.005055	\\
3.114	0.003886	\\
3.884	0.003471	\\
4.843	0.003402	\\
6.04	0.003381	\\
7.532	0.003645	\\
9.393	0.003793	\\
11.71	0.003517	\\
14.61	0.003302	\\
18.22	0.003316	\\
22.72	0.003122	\\
28.33	0.003235	\\
35.33	0.003371	\\
44.07	0.003461	\\
54.95	0.003518	\\
68.53	0.003485	\\
85.47	0.003388	\\
106.6	0.003442	\\
132.9	0.003422	\\
165.8	0.003368	\\
206.7	0.003234	\\
257.8	0.003201	\\
321.5	0.003193	\\
400.9	0.003228	\\
500	0.003224	\\
};		
\addlegendentry{PEO 1M 2000 ppm}

\addplot[color=brown, mark=star, mark size=2.0pt, line width=0.5pt]		
table[row sep=crcr]{		
1.032	0.001393	\\
1.287	0.001333	\\
1.606	0.001847	\\
2.002	0.002508	\\
2.497	0.0026	\\
3.114	0.002155	\\
3.884	0.001922	\\
4.843	0.001618	\\
6.04	0.002336	\\
7.532	0.002256	\\
9.393	0.00221	\\
11.71	0.001934	\\
14.61	0.001583	\\
18.22	0.001726	\\
22.72	0.001621	\\
28.33	0.001546	\\
35.33	0.001712	\\
44.07	0.001941	\\
54.95	0.002065	\\
68.53	0.002042	\\
85.47	0.001934	\\
106.6	0.002022	\\
132.9	0.002014	\\
165.8	0.001991	\\
206.7	0.00192	\\
257.8	0.001872	\\
321.5	0.001876	\\
400.9	0.001884	\\
500	0.001969	\\
};		
\addlegendentry{PEO 1M 1000 ppm}

\addplot[color=teal, mark=triangle*, mark size=2.0pt, line width=0.5pt]		
table[row sep=crcr]{		
0.1	0.02345	\\
0.1207	0.03447	\\
0.1456	0.03491	\\
0.1758	0.02734	\\
0.2121	0.02017	\\
0.256	0.01687	\\
0.3089	0.01452	\\
0.3728	0.01106	\\
0.4498	0.009003	\\
0.5429	0.00736	\\
0.6551	0.009921	\\
0.7906	0.007761	\\
0.9541	0.004703	\\
1.151	0.002891	\\
1.389	0.002663	\\
1.677	0.002166	\\
2.024	0.002489	\\
2.442	0.002021	\\
2.947	0.00203	\\
3.556	0.001652	\\
4.292	0.001521	\\
5.179	0.001582	\\
6.251	0.001645	\\
7.543	0.001697	\\
9.103	0.001578	\\
10.99	0.001523	\\
13.26	0.00149	\\
16	    0.001406	\\
19.31	0.00125	\\
23.3	0.001361	\\
28.12	0.001304	\\
33.93	0.001311	\\
40.95	0.001452	\\
49.42	0.001442	\\
59.64	0.001462	\\
71.97	0.001454	\\
86.85	0.001418	\\
104.8	0.001443	\\
126.5	0.001456	\\
152.6	0.001435	\\
184.2	0.001413	\\
222.3	0.001381	\\
268.3	0.001333	\\
323.7	0.001327	\\
390.7	0.001331	\\
471.5	0.001431	\\
500	0.001427	\\
};		
\addlegendentry{PEO 1M 500 ppm}

\end{axis}
\end{tikzpicture}
    \caption{Viscosity $\mu$ as a function of shear rate $\dot{\gamma}$ of PEO-1M.}
    \label{fig:viscosity_PEO}
\end{figure}

\begin{figure}[h] 
    \centering
    \pgfplotsset{compat=1.17} 

\definecolor{mycolor0}{RGB}{50,155,105}    
\definecolor{mycolor5}{RGB}{213,81,156}    
\definecolor{mycolor10}{RGB}{54,80,163}    
\definecolor{mycolor125}{RGB}{244,196,48}  
\definecolor{mycolor15}{RGB}{245,127,34}   
\definecolor{mycolorSplash}{RGB}{204,204,255} 
\definecolor{mycolorNo}{RGB}{224,224,224}     

\begin{tikzpicture}
\begin{axis}[
width=1.2\figW,
height=1.0\figH,
at={(0\figW,0\figH)},
    xlabel={\small Shear Rate (s$^{-1}$)},
    ylabel={\small N1 (Pa)},
    xmin=1, xmax=10000,
    ymin=0.1, ymax=10000,
    xmode=log,
    ymode=log,
    legend style={at={(1.10,0.90)}, anchor=north west, draw=none, fill=none, font=\footnotesize},
    ticklabel style = {font=\footnotesize}, legend cell align=left, legend image post style={scale=0.8}, 
    row sep=-2pt 
    ]

\addplot[color=teal, mark=triangle*, mark size=2.0pt, line width=0.5pt]
table[row sep=crcr]{
4	11.06938645	\\
6	17.96304526	\\
10	36.18864367	\\
20	53.57154354	\\
30	76.131999	\\
40	103.9937985	\\
60	154.1754349	\\
100	268.4423065	\\
200	560.271179	\\
300	891.8308229	\\
400	1335.895014	\\
};
\addlegendentry{10000 ppm}

\addplot[color=red, mark=square*, mark size=2.0pt, line width=0.5pt]
table[row sep=crcr]{
60	5.217929111	\\
100	8.138228889	\\
200	23.34285267	\\
300	40.81967811	\\
400	62.72263456	\\
500	87.18911167	\\
600	113.465535	\\
700	142.3709826	\\
800	354.7755549	\\
900	472.1794471	\\
};
\addlegendentry{2500 ppm}

\addplot[color=orange, mark=*, mark size=2.0pt, line width=0.5pt]
table[row sep=crcr]{
200	4.796192	\\
300	8.334957	\\
400	12.321999	\\
500	21.277386	\\
600	32.657723	\\
700	58.49094	\\
};
\addlegendentry{1000 ppm}

\addplot[color=violet, mark=triangle, mark size=2.0pt, line width=0.5pt]
table[row sep=crcr]{
300	6.879898	\\
400	10.115756	\\
500	13.207574	\\
600	17.039109	\\
700	29.533368	\\
800	40.830023	\\
900	49.090085	\\
1000	51.216972	\\
1400	157.359658	\\
2000	144.5427513	\\
};
\addlegendentry{500 ppm}

\addplot[color=blue, mark=diamond*, mark size=2.0pt, line width=0.5pt]
table[row sep=crcr]{
600	6.57234125	\\
700	7.1101125	\\
800	11.4838	\\
900	16.43795708	\\
1000	19.70126792	\\
2000	59.84152125	\\
};
\addlegendentry{250 ppm}

\addplot[color=pink, mark=diamond, mark size=2.0pt, line width=0.5pt]
table[row sep=crcr]{
500	2.071201	\\
800	4.385872	\\
1000	9.383755	\\
1500	26.590939	\\
2000	46.06322	\\
2500	60.385618	\\
3000	65.90761	\\
4000	79.008493	\\
5000	84.324415	\\
6000	83.047966	\\
};
\addlegendentry{150 ppm}

\addplot[color=black, mark=x, mark size=2.0pt, line width=0.5pt]
table[row sep=crcr]{
800	0.649361333	\\
1000	2.124191333	\\
1500	12.613184	\\
2000	28.385348	\\
2500	43.053986	\\
3000	49.79561933	\\
4000	91.53015	\\
5000	90.31691933	\\
6000	130.3957607	\\
};
\addlegendentry{100 ppm}

\addplot[color=brown, mark=star, mark size=2.0pt, line width=0.5pt]
table[row sep=crcr]{
500	0.408415	\\
800	1.384542	\\
1000	1.60728	\\
1500	4.959782	\\
2000	13.327594	\\
2500	19.306355	\\
};
\addlegendentry{50 ppm}

\addplot[color=magenta, mark=triangle, mark size=2.0pt, line width=0.5pt]
table[row sep=crcr]{
800	0.43572	\\
1000	-0.376469	\\
1500	1.000111	\\
2000	3.305792	\\
2500	4.356869	\\
3000	7.317033	\\
4000	12.374424	\\
5000	20.673449	\\
6000	27.968923	\\
};
\addlegendentry{25 ppm}

\end{axis}
\end{tikzpicture}
    \caption{First normal stress difference $N_1$ as a function of shear rate $\dot{\gamma}$ of PEO-8M at various concentrations.}
    \label{fig:n1_peo_4m}
\end{figure}

\subsubsection{Extensional Rheometry Relaxation Time}

To evaluate the extensional viscoelastic behavior of the samples, Capillary Breakup Extensional Rheometry (CaBER) was utilized \cite{anna2001elasto}, adhering to the experimental procedures outlined by \cite{gaillard2022determines} and \cite{varkevisser2026effect}. The protocol involves confining a small droplet of the polymer solution between two parallel circular plates ($5~\text{mm}$ in diameter). These plates are rapidly pulled apart to generate an unstable liquid bridge. The subsequent temporal evolution of the thinning fluid filament is captured via high-speed imaging. Following an early inertio-capillary necking phase, the dynamics of the polymer solutions transition into an elasto-capillary balance. Governed by the Oldroyd-B constitutive model, the minimum diameter of the filament in this regime undergoes a characteristic exponential decay \cite{bazilevskii1997failure,clasen2006dilute}:
\begin{equation}
    D_{\text{min}} \propto \exp \left( -\frac{t}{3\tau} \right).
\end{equation}
The extensional relaxation times ($\tau$) extracted from these exponential fits are compiled in Table~\ref{tab:fluid_matrix}.

\subsection{Theoretical Polymeric Relaxation}

To evaluate the elastic capacity of the polymer solutions independently from the CaBER extensional data, we calculate the theoretical entropic modulus ($G_Z$). Following the framework validated by Clasen \textit{et al.} \cite{clasen2006dilute} for dilute polymer solutions, the modulus is dictated by the polymer chain density:
\begin{equation}
    G_Z = n k_B T = \frac{c N_A k_B T}{M_w}
\end{equation}
where $c$ is the polymer concentration, $N_A$ is Avogadro's constant, $k_B$ is the Boltzmann constant, $T$ is the absolute temperature, and $M_w$ is the molecular weight. While the highest concentrations in our quasi-Boger regime approach the polymer overlap concentration ($c^*$), this dilute-limit approximation provides a conservative, leading-order bound for the structural elastic capacity.

This theoretical elastic modulus is a material property that is characterized by the polymeric contribution to the viscosity ($\mu_p = \mu_0 - \mu_s$), where $\mu_s$ represents the solvent viscosity, rather than the total bulk viscosity ($\mu_0$). Thus, the theoretical Zimm relaxation time ($\tau_Z$) is defined as \cite{gaillard2022determines}:
\begin{equation}
    \tau_Z = \frac{\mu_p}{G_Z}
\end{equation}

During droplet impact, the extreme deformation rates impose a characteristic bulk shear rate ($\dot\gamma_s$). The relative strength of the elastic forces driving the polymer coil-stretch transition is subsequently parameterized by the dynamic Weissenberg number:
\begin{equation}
    \Wi = 2 \dot\gamma_s \tau_Z.
\end{equation}

\subsection{Materials}

Aqueous polymer solutions were prepared using polyethylene oxide with molecular weights (PEO 1M, 2M, and 8M) and polyacrylamide (PAAM 5M and 15M). A stock solution of 10,000 ppm (weight parts per million) was produced by mixing for 16 h with a mechanical stirrer until a homogeneous solution was obtained. To prevent aggregation of the polymer upon contact with water, the polymer powder was mixed with an equal mass of isopropanol before adding water. This protocol ensured highly reproducible solutions as confirmed by rheological measurements. The rheological properties of the solutions used in this study closely matched those described in detail by \cite{gaillard2024beware,gaillard2022determines,varkevisser2026effect}.

\subsection{Fluid Matrix}

The complete experimental matrix, which captures the zero-shear viscosity ($\mu_0$) and the characteristic relaxation time ($\tau$) for each formulation. To systematically decouple the effects of elasticity from variable viscosity during impact, the solutions are categorized into three distinct rheological regimes:

\begin{enumerate}
    \item \textbf{Boger Fluids:} PAAM in high-viscosity (60, 70, 80, and 90 vol.\%) aqueous glycerol solutions. These fluids possess a constant high-shear viscosity and high elasticity.
    \item \textbf{Quasi-Boger Fluids:} Dilute PEO and PAAM (concentration $\leq 1000$ ppm) in DI water, which have minimal shear-thinning (thinning exponent $n \approx 0.95 - 1.0$), yet possess high elasticity. 
    \item \textbf{Shear Thinning:} Concentrated PEO and PAAM (concentration $2000-10000$ ppm) in purely aqueous solvents. 
\end{enumerate}

By comparing the spreading dynamics across these three specific regimes, we can experimentally isolate the mechanical influence of elastic stresses from the kinematic enhancements induced by shear thinning.

\begin{table*}[h]
    \centering
    \caption{Summary of the experimental fluid matrix for the Boger fluids.}
    \vspace{0.2cm}
    \begin{tabular}{@{}llcccc@{}}
        \toprule
        \textbf{Polymer} & \textbf{$M_w$ (g/mol)} & \textbf{Conc. (ppm)} & \textbf{Solvent} & \textbf{$\mu_0$ (Pa$\cdot$s)} & \textbf{$\tau$ (ms)} \\
        \midrule
        \multicolumn{6}{c}{\textbf{Constant-Viscosity Boger Fluids}} \\
        \midrule
        PAAM & $5 \times 10^6$& 1000 & 90\% Glycerol (aq) & 0.1933 & 28.1 \\
             &                & 1000 & 80\% Glycerol (aq) & 0.0741 & 22.5 \\
             &                & 1000 & 70\% Glycerol (aq) & 0.0327 & 13.6 \\
             &                & 1000 & 60\% Glycerol (aq) & 0.0158 & 10.2 \\
             &                & 1000 & Water              & 0.0019 & 6.5 \\
        \bottomrule
    \end{tabular}
    \label{tab:fluid_matrix_boger}
\end{table*}

\begin{table*}[h]
    \centering
    \caption{Summary of the experimental fluid matrix for the Quasi-Boger fluids.}
    \vspace{0.2cm}
    \begin{tabular}{@{}llcccc@{}}
        \toprule
        \textbf{Polymer} & \textbf{$M_w$ (g/mol)} & \textbf{Conc. (ppm)} & \textbf{Solvent} & \textbf{$\mu_0$ (Pa$\cdot$s)} & \textbf{$\tau$ (ms)} \\
        \midrule
        \multicolumn{6}{c}{\textbf{Quasi-Boger Fluids}} \\
        \midrule
        PEO & $1 \times 10^6$ & 1000  & Water & 0.0019 & 3.24 \\
            &                 & 500   & Water & 0.0014 & 2.39 \\
            &                 & 350   & Water & 0.0013 & 2.19 \\
            &                 & 200   & Water & 0.0012 & 1.69 \\
            &                 & 100   & Water & 0.0011 & 0.82 \\
            &                 & 50    & Water & 0.0011 & 0.36 \\
        \midrule
        PEO & $2 \times 10^6$ & 1000  & Water & 0.0022 & 3.59 \\
            &                 & 500   & Water & 0.0016 & 2.80 \\
            &                 & 350   & Water & 0.0014 & 2.38 \\
            &                 & 200   & Water & 0.0012 & 1.71 \\
            &                 & 100   & Water & 0.0011 & 1.10 \\
            &                 & 50    & Water & 0.0011 & 0.46 \\
        \midrule
        PEO & $8 \times 10^6$ & 800   & Water & 0.0068 & 103.69 \\
            &                 & 500   & Water & 0.0038 & 85.28 \\
            &                 & 350   & Water & 0.0027 & 62.94 \\
            &                 & 200   & Water & 0.0017 & 41.43 \\
            &                 & 100   & Water & 0.0013 & 26.55 \\
            &                 & 50    & Water & 0.0012 & 14.44 \\
            &                 & 20    & Water & 0.0011 & 6.81 \\
        \midrule
        PAAM & $15 \times 10^6$& 1000  & Water & 0.0029 & 16.82 \\
             &                 & 500   & Water & 0.0019 & 10.24 \\
             &                 & 200   & Water & 0.0013 & 6.50 \\
             &                 & 100   & Water & 0.0011 & 2.51 \\
             &                 & 50    & Water & 0.0011 & 1.38 \\
        \bottomrule
    \end{tabular}
    \label{tab:fluid_matrix_quasi}
\end{table*}        
        
\begin{table*}[h]
    \centering
    \caption{Summary of the experimental fluid matrix for the shear-thinning viscoelastic fluids.}
    \vspace{0.2cm}
    \begin{tabular}{@{}llcccc@{}}
        \toprule
        \textbf{Polymer} & \textbf{$M_w$ (g/mol)} & \textbf{Conc. (ppm)} & \textbf{Solvent} & \textbf{$\mu_0$ (Pa$\cdot$s)} & \textbf{$\tau$ (ms)} \\
        \midrule
        \multicolumn{6}{c}{\textbf{Shear-Thinning Viscoelastic Fluids}} \\
        \midrule
        PEO & $1 \times 10^6$ & 10000 & Water & 0.0580 & 18.05 \\
            &                 & 7500  & Water & 0.0275 & 13.93 \\
            &                 & 5000  & Water & 0.0115 & 10.20 \\
            &                 & 3500  & Water & 0.0066 & 7.41 \\
            &                 & 2000  & Water & 0.0033 & 4.87 \\
        \midrule
        PEO & $2 \times 10^6$ & 10000 & Water & 0.0980 & 17.83 \\
            &                 & 5000  & Water & 0.0166 & 10.00 \\
            &                 & 2000  & Water & 0.0042 & 4.96 \\
        \midrule
        PAAM & $15 \times 10^6$& 10000 & Water & 0.4300 & 146.78 \\
             &                 & 5000  & Water & 0.0370 & 66.20 \\
             &                 & 2000  & Water & 0.0069 & 34.14 \\
        \bottomrule
    \end{tabular}
    \label{tab:fluid_matrix_thinning}
\end{table*}

\subsection{Interfacial Properties}

Figure \ref{fig:surface tension} presents the evolution of surface tension, $\sigma$, over time, measured via maximum bubble pressure tensiometry (data from \cite{varkevisser2026effect}). While the presence of macromolecules typically lowers the equilibrium surface tension of an aqueous solution at long timescales, the dynamic measurements reveal that at early timescales ($t \le 10^{-2}$ s), $\sigma$ converges toward that of pure water. Extrapolating these adsorption dynamics into the impact regime using the empirical Hua-Rosen model \cite{hua1988dynamic} confirms that $\sigma$ remains similar to the pure solvent plateau ($\approx 72~\text{mN m}^{-1}$) throughout the characteristic timescale of droplet impact ($10^{-4}$ to $10^{-3}$ s). Moreover, the continuous and rapid creation of a fresh liquid-air interface outpaces the diffusion and adsorption timescales of the polymer chains \cite{varghese2024effect}. Consequently, the relevant surface tension governing the impact dynamics is unaffected by the polymer, justifying the use of the constant solvent surface tension in all subsequent calculations of this work. 

\begin{figure}[h]
    \centering
    \pgfplotsset{compat=1.17} 

\definecolor{mycolor0}{RGB}{50,155,105}    
\definecolor{mycolor5}{RGB}{213,81,156}    
\definecolor{mycolor10}{RGB}{54,80,163}    
\definecolor{mycolor125}{RGB}{244,196,48}  
\definecolor{mycolor15}{RGB}{245,127,34}   
\definecolor{mycolorSplash}{RGB}{204,204,255} 
\definecolor{mycolorNo}{RGB}{224,224,224}     

\begin{tikzpicture}
\begin{axis}[
width=1.4\figW,
height=1.0\figH,
at={(0\figW,0\figH)},
    xlabel={\small $t$ (s)},
    ylabel={\small $\sigma$},
    xmin=0.00001, xmax=100,
    ymin=55, ymax=80,
    xmode=log,
    ytick={55, 60, 65 , 70, 75, 80},
    legend style={at={(0.02,0.50)}, anchor=north west, draw=none, fill=none, font=\footnotesize},
    ticklabel style = {font=\footnotesize}, legend cell align=left, legend image post style={scale=0.8}, 
    row sep=-2pt 
    ]

\fill[gray!20] (0.0006, 50) rectangle (0.004, 80);

\draw[<->, thick, >=stealth] (0.0006, 73.5) -- (0.004, 73.5) 
    node[midway, above, align=center, font=\footnotesize] {$\tau_{\text{impact}}$};

\addplot[  line width=2.0pt,  domain=0.00001:100, samples=200] {59 +((72 - 59 )/(1+(x/0.3)^0.8))};
\addlegendentry{Hua-Rosen Eq.}

\addplot[only marks, color=red, mark=o, mark size=2.5pt, line width=0.5pt]
table[row sep=crcr]{
0.026608302	71.80783444	\\
0.036603347	72.01657718	\\
0.048116862	72.08357232	\\
0.065229586	71.82957847	\\
0.088382499	72.00294308	\\
0.116213242	71.85625899	\\
0.152741446	72.06570695	\\
0.207045888	71.88293951	\\
0.289106725	72.30512642	\\
0.352401412	71.98155165	\\
0.485028146	71.76293594	\\
0.63773092	71.65186505	\\
0.877551773	71.61131536	\\
1.13608508	71.74977198	\\
};
\addlegendentry{water}

\addplot[only marks, color=teal, mark=o, mark size=2.5pt, line width=0.5pt]
table[row sep=crcr]{
0.027022827	71.59392013	\\
0.034980924	71.80360316	\\
0.048125197	71.9411195	\\
0.064237321	71.93665316	\\
0.08574372	71.93218682	\\
0.116233372	71.71380617	\\
0.157544389	71.60226514	\\
0.207099687	71.66926028	\\
0.26413722	71.59427274	\\
0.363403958	71.69617587	\\
0.492669619	71.40656881	\\
0.64766599	71.43795075	\\
0.877817809	71.36202292	\\
1.171859929	71.25071697	\\
1.564397167	71.13941101	\\
2.088603246	70.95687864	\\
2.704623962	70.88165603	\\
3.557206387	70.5212927	\\
4.972238861	70.08876268	\\
6.343845349	69.7288695	\\
8.47397509	69.11897866	\\
12.21997094	68.04494081	\\
15.60979988	66.68787788	\\
};
\addlegendentry{50 ppm}

\addplot[only marks, color=brown, mark=o, mark size=2.5pt, line width=0.5pt]
table[row sep=crcr]{
0.026233057	70.98896578	\\
0.036082459	71.30454815	\\
0.048158551	71.37130822	\\
0.065269142	71.3309936	\\
0.085814293	71.25553591	\\
0.116293785	71.28644771	\\
0.155268816	71.06830213	\\
0.204117175	71.09968407	\\
0.276663362	70.98814304	\\
0.37491233	71.05466803	\\
0.471077033	70.76623633	\\
0.658240207	70.61861195	\\
0.839563154	70.50801121	\\
1.155633928	70.21816908	\\
1.519991858	69.82219255	\\
2.124726269	69.35404933	\\
2.75318893	68.74462864	\\
3.622805755	67.99252005	\\
4.91572174	66.99064889	\\
6.672078893	65.73948529	\\
8.525502271	64.13312992	\\
15.94385366	61.77302047	\\
};
\addlegendentry{100 ppm}

		

\addplot[only marks, color=blue, mark=square, mark size=2.5pt, line width=0.5pt]
table[row sep=crcr]{
0.027066156	70.27623154	\\
0.036701748	69.80855845	\\
0.048292198	69.09206307	\\
0.065470115	68.80245602	\\
0.08878526	68.26355653	\\
0.113291647	67.79682373	\\
0.155895352	67.75627404	\\
0.208133551	67.57374167	\\
0.273648972	67.49828398	\\
0.354713153	66.60395765	\\
0.473613237	66.35019887	\\
0.652168768	65.7398379	\\
0.871681586	64.63136219	\\
1.146959546	63.91486681	\\
1.533743861	62.41464584	\\
2.079848974	61.91135955	\\
2.73643047	61.26609059	\\
3.655741092	60.54936014	\\
4.882204852	60.11753533	\\
6.617410845	60.0059943	\\
8.834041101	59.89468834	\\
11.61478413	59.81923066	\\
15.74284586	59.70768963	\\
};
\addlegendentry{200 ppm}

\end{axis}
\end{tikzpicture}
    \caption{Dynamic surface tension as a function of time for pure water and various PEO-4M polymer solutions (data from \cite{varkevisser2026effect}). The solid line represents the Hua-Rosen adsorption model \cite{hua1988dynamic} fitted to the highly concentrated 200 ppm data.}
    \label{fig:surface tension}
\end{figure}

\clearpage

\section{Derivation of the Elasto-Viscocapillary Closure}

\subsection{Total Available Energy Density ($\mathcal{E}$)}
The initial state of the droplet is defined by its kinetic energy ($E_k$) and initial surface energy ($E_{s,0}$). We define the initial radius $R_0 = D_0/2$ and volume $\Omega = \frac{4}{3}\pi R_0^3 = \frac{\pi}{6}D_0^3$.
\begin{equation}
    E_k = \frac{1}{2}\rho \Omega V_0^2 = \frac{1}{2}\rho \left( \frac{\pi}{6}D_0^3 \right) V_0^2
\end{equation}
\begin{equation}
    E_{s,0} = \pi \sigma D_0^2
\end{equation}
The total initial energy is the sum of these components:
\begin{equation}
    E_{total} = E_k + E_{s,0} = \frac{1}{2}\rho \left( \frac{\pi}{6}D_0^3 \right) V_0^2 + \pi \sigma D_0^2
\end{equation}
To non-dimensionalize, we divide the total energy by the kinetic energy:
\begin{equation}
    \mathcal{E} = \frac{E_k + E_{s,0}}{E_k} = 1 + \frac{\pi \sigma D_0^2}{\frac{1}{2}\rho \left( \frac{\pi}{6}D_0^3 \right) V_0^2} = 1 + \frac{12 \sigma}{\rho D_0 V_0^2}
\end{equation}
Substituting the Weber number ($\We= \rho D_0 V_0^2 / \sigma$):
\begin{equation}
    \mathcal{E} = 1 + \frac{12}{We}
\end{equation}

\subsection{Final Surface Energy Term}
At maximum spreading, the fluid expands to a radius $a_{\text{max}} = \beta_{\text{max}} R_0$. The geometric scaling of the final surface energy ($E_{s,max}$) is:
\begin{equation}
    E_{s,max} \sim \sigma D_{\text{max}}^2
\end{equation}
Applying the kinematic relation $D_{\text{max}} = V_s t_s$:
\begin{equation}
    E_{s,max} \sim \sigma V_s^2 t_s^2
\end{equation}
Normalizing by the initial kinetic energy:
\begin{equation}
    \frac{E_{s,max}}{E_k} \sim \frac{\sigma V_s^2 t_s^2}{\rho D_0^3 V_0^2}
\end{equation}
We define the dimensionless velocity $V = V_s / V_0$ and dimensionless time $T = t_s / \tau_{ic}$, where $\tau_{ic} = \sqrt{\rho D_0^3 / \sigma}$. Substituting $V_s = V V_0$ and $t_s = T \sqrt{\rho D_0^3 / \sigma}$:
\begin{equation}
    \frac{E_{s,max}}{E_k} \sim \frac{\sigma (V V_0)^2 \left( T \sqrt{\frac{\rho D_0^3}{\sigma}} \right)^2}{\rho D_0^3 V_0^2} = \frac{\sigma V^2 V_0^2 T^2 \left( \frac{\rho D_0^3}{\sigma} \right)}{\rho D_0^3 V_0^2}
\end{equation}
Canceling the dimensional parameters yields the dimensionless surface energy sink:
\begin{equation}
    \frac{E_{s,max}}{E_k} \sim V^2 T^2
\end{equation}

\subsection{Viscous Dissipation Term}
The instantaneous viscous dissipation rate ($\dot{W}$) is defined by the volume-limited lubrication flow, where $h$ is the film thickness:
\begin{equation}
    \dot{W} \sim \mu \left( \frac{V_s}{h} \right)^2 D^2 h = \mu \frac{V_s^2}{h} D^2
\end{equation}
Applying volume conservation $\Omega \sim D_0^3 \sim D^2 h$, we isolate $h \sim D_0^3/D^2$ and substitute it into the dissipation rate:
\begin{equation}
    \dot{W} \sim \mu \frac{V_s^2}{\left( \frac{D_0^3}{D^2} \right)} D^2 = \mu \frac{V_s^2 D^4}{D_0^3}
\end{equation}
Substituting the instantaneous diameter $D(t) = V_s t$:
\begin{equation}
    \dot{W}(t) \sim \mu \frac{V_s^2 (V_s t)^4}{D_0^3} = \mu \frac{V_s^6 t^4}{D_0^3}
\end{equation}
The total viscous work ($W_v$) is the integral over the spreading time ($t_s$):
\begin{equation}
    W_v = \int_0^{t_s} \dot{W}(t) dt \sim \int_0^{t_s} \mu \frac{V_s^6 t^4}{D_0^3} dt \sim \mu \frac{V_s^6 t_s^5}{D_0^3}
\end{equation}
Normalizing by the initial kinetic energy:
\begin{equation}
    \frac{W_v}{E_k} \sim \frac{\mu V_s^6 t_s^5}{\rho D_0^6 V_0^2}
\end{equation}
Substituting the dimensionless variables $V_s = V V_0$ and $t_s = T \sqrt{\rho D_0^3 / \sigma}$:
\begin{equation}
    \frac{W_v}{E_k} \sim \frac{\mu (V V_0)^6 \left( T \sqrt{\frac{\rho D_0^3}{\sigma}} \right)^5}{\rho D_0^6 V_0^2} = \left( \frac{\mu \rho^{3/2} D_0^{3/2} V_0^4}{\sigma^{5/2}} \right) V^6 T^5
\end{equation}
Factoring the dimensional coefficient into standard dimensionless numbers:
\begin{equation}
    \frac{\mu \rho^{3/2} D_0^{3/2} V_0^4}{\sigma^{5/2}} = \left( \frac{\rho^2 D_0^2 V_0^4}{\sigma^2} \right) \left( \frac{\mu}{\sqrt{\rho \sigma D_0}} \right) = \We^2 \Oh
\end{equation}
Defining the viscous damping parameter $\Lambda_{\nu}= \We^2 \Oh$:
\begin{equation}
    \frac{W_v}{E_k} \sim \Lambda_{\nu}V^6 T^5
\end{equation}

\subsection{Elastic Strain Energy Term}
The stored elastic strain energy ($E_{el}$) is derived from the integral of the Hertzian contact force. For an incompressible solid ($\nu = 0.5$), the stiffness is $K = \frac{8}{3} G \sqrt{R_0}$:
\begin{equation}
    E_{el} = \frac{2}{5} K \delta_{\text{max}}^{5/2} = \frac{2}{5} \left( \frac{8}{3} G \sqrt{R_0} \right) \delta_{\text{max}}^{5/2}
\end{equation}
Substituting the geometric strain condition $\delta_{\text{max}} = \beta_{\text{max}}^2 R_0$:
\begin{equation}
    E_{el} = \frac{16}{15} G \sqrt{R_0} \left( \beta_{\text{max}}^2 R_0 \right)^{5/2} = \frac{16}{15} G R_0^3 \beta_{\text{max}}^5
\end{equation}
Substituting the droplet volume $R_0^3 = \frac{3\Omega}{4\pi}$:
\begin{equation}
    E_{el} = \frac{16}{15} G \left( \frac{3\Omega}{4\pi} \right) \beta_{\text{max}}^5 = \frac{4}{5\pi} G \Omega \beta_{\text{max}}^5
\end{equation}
Normalizing by the initial kinetic energy:
\begin{equation}
    \frac{E_{el}}{E_k} = \frac{\frac{4}{5\pi} G \Omega \beta_{\text{max}}^5}{\frac{1}{2}\rho \Omega V_0^2} = \frac{8}{5\pi} \left( \frac{G}{\rho V_0^2} \right) \beta_{\text{max}}^5 
\end{equation}
We define the kinematic link by non-dimensionalizing $\beta_{\text{max}} = D_{\text{max}} / D_0$:
\begin{equation}
    \beta_{\text{max}} = \frac{V_s t_s}{D_0} = \frac{V V_0 T \sqrt{\frac{\rho D_0^3}{\sigma}}}{D_0} = V T \sqrt{\We}
\end{equation}
Substituting $\beta_{\text{max}}$ into the normalized elastic energy, where $\El = \frac{G}{\rho V_0^2}$:
\begin{equation}
    \frac{E_{el}}{E_k} = \frac{8}{5\pi} \El \left( V T \sqrt{\We} \right)^5 = \left( \frac{8}{5\pi} \El \We^{5/2} \right) V^5 T^5
\end{equation}
Defining the elastic damping parameter: 
\begin{equation}
    \Lambda_{el} = \frac{8}{5\pi} \El \We^{5/2}
\end{equation}
then the contribution of elastic energy becomes:
\begin{equation}
    \frac{E_{el}}{E_k} = \Lambda_{el} V^5 T^5
\end{equation}

\subsection{Dimensionless Energy Polynomial}
Equating the total available energy to the energy sinks yields the complete balance:
\begin{equation}
\boxed{
    \mathcal{E} \approx V^2 T^2 + \Lambda_{\nu}V^6 T^5 + \Lambda_{el} V^5 T^5}
\end{equation}

\subsection{Derivation of the Spreading Time ($T$)}
Factoring $V^6 T^5$ out of the damping sinks:
\begin{equation}
    \mathcal{E} \approx V^2 T^2 + \left[ \Lambda_{\nu}+ \Lambda_{el} V^{-1} \right] V^6 T^5
\end{equation}
Substituting the leading-order velocity closure $V \approx \mathcal{E}^{1/2}$ into the damping bracket defines the effective damping parameter $\Lambda_{eff}$ (further justification on this assumption is in \cite{abbot2026kinematic} and copied at the end of this section) :
\begin{equation}
\boxed{
    \Lambda_{eff} = \Lambda_{\nu}+ \Lambda_{el} \mathcal{E}^{-1/2}
    }
\end{equation}
\begin{equation}
    \mathcal{E} \approx V^2 T^2 + \Lambda_{eff} V^6 T^5
\end{equation}
Extracting the asymptotic limits for $T$:
\begin{equation}
    \text{Inertial Limit } (\Lambda_{eff} \to 0): \quad \mathcal{E} \approx V^2 T^2 \implies T_{in} \approx \frac{\mathcal{E}^{1/2}}{V}
\end{equation}
\begin{equation}
    \text{Damping Limit } (\Lambda_{eff} \gg 1): \quad \mathcal{E} \approx \Lambda_{eff} V^6 T^5 \implies T_{damp} \approx \left( \frac{\mathcal{E}}{\Lambda_{eff} V^6} \right)^{1/5}
\end{equation}
Applying the harmonic mean interpolation ($T^{-1} \approx T_{in}^{-1} + T_{damp}^{-1}$):
\begin{equation}
    T \approx \left[ \frac{V}{\mathcal{E}^{1/2}} + \left( \frac{\Lambda_{eff} V^6}{\mathcal{E}} \right)^{1/5} \right]^{-1}
\end{equation}
Factoring out the inertial root:
\begin{equation}
    T \approx \frac{\mathcal{E}^{1/2}}{V} \left[ 1 + \left( \Lambda_{eff} V \mathcal{E}^{3/2} \right)^{1/5} \right]^{-1}
\end{equation}
Decoupling time from velocity by substituting the geometric prefactor $T_0 = \sqrt{\pi/6}$ for $\mathcal{E}^{1/2}/V$, and substituting the leading-order closure $V \approx \mathcal{E}^{1/2}$ (further justification on this assumption is in \cite{abbot2026kinematic} and copied at the end of this section) into the exponent:
\begin{equation}
    T \approx  T_0 \left[ 1 + (\Lambda_{eff} \mathcal{E}^2)^{1/5} \right]^{-1}
\end{equation}
Expanding $\Lambda_{eff}$:
\begin{equation}
    \Lambda_{eff} \mathcal{E}^2 = \left( \Lambda_{\nu}+ \Lambda_{el} \mathcal{E}^{-1/2} \right) \mathcal{E}^2 = \Lambda_{\nu}\mathcal{E}^2 + \Lambda_{el} \mathcal{E}^{3/2}
\end{equation}
\begin{equation}
\boxed{
    T \approx T_0 \left[ 1 + \left( \Lambda_{\nu}\mathcal{E}^2 + \Lambda_{el} \mathcal{E}^{3/2} \right)^{1/5} \right]^{-1}}
\end{equation}

\subsection{Derivation of the Spreading Velocity ($V$)}
Returning to the reduced polynomial:
\begin{equation}
    \mathcal{E} \approx V^2 T^2 + \Lambda_{eff} V^6 T^5
\end{equation}
Extracting the asymptotic limits for $V$:

\textbf{Inertial Limit} ($\Lambda_{eff} \to 0$):
\begin{equation}
     \quad V_{in} \approx \frac{\mathcal{E}^{1/2}}{T}
\end{equation}

\textbf{Damping Limit} ($\Lambda_{eff} \gg 1$):
\begin{equation}
     \quad V_{damp} \approx \left( \frac{\mathcal{E}}{\Lambda_{eff} T^5} \right)^{1/6}
\end{equation}
Applying a 6th-power harmonic mean interpolation ($V^{-6} \approx V_{in}^{-6} + V_{damp}^{-6}$):
\begin{equation}
    V^{-6} \approx \left( \frac{T}{\mathcal{E}^{1/2}} \right)^6 + \frac{\Lambda_{eff} T^5}{\mathcal{E}} = \frac{T^6}{\mathcal{E}^3} + \frac{\Lambda_{eff} T^5}{\mathcal{E}}
\end{equation}
\begin{equation}
    V \approx \left[ \frac{T^6}{\mathcal{E}^3} + \frac{\Lambda_{eff} T^5}{\mathcal{E}} \right]^{-1/6}
\end{equation}
Factoring $\mathcal{E}^{-3}$ from the bracket:
\begin{equation}
    V \approx \left( \mathcal{E}^{-3} \right)^{-1/6} \left[ T^6 + \Lambda_{eff} \mathcal{E}^2 T^5 \right]^{-1/6}
\end{equation}
\begin{equation}
\boxed{
    V \approx \mathcal{E}^{1/2} \left[ T^6 + \Lambda_{eff} \mathcal{E}^2 T^5 \right]^{-1/6}}
\end{equation}

\subsection{Derivation of the Maximum Spreading Ratio ($\beta_{\text{max}}$)}
The maximum spreading ratio is defined by the dimensionless kinematic link:
\begin{equation}
    \beta_{\text{max}} = V T \sqrt{We}
\end{equation}
Applying the zeroth-order closure $V \approx \mathcal{E}^{1/2}$ for the spreading velocity specifically within the evaluation of the temporal exponent:
\begin{equation}
    \beta_{\text{max}} \approx \mathcal{E}^{1/2} \sqrt{We} \left[ 1 + (\Lambda_{eff} \mathcal{E}^2)^{1/5} \right]^{-1}
\end{equation}
\begin{equation}
    \beta_{\text{max}} \approx \sqrt{\We\mathcal{E}} \left[ 1 + (\Lambda_{eff} \mathcal{E}^2)^{1/5} \right]^{-1}
\end{equation}
Expanding the effective damping parameter yields the final analytical closure:
\begin{equation}
    \beta_{\text{max}} \approx \sqrt{\We\mathcal{E}} \left[ 1 + \left( \Lambda_{\nu}\mathcal{E}^2 + \Lambda_{el} \mathcal{E}^{3/2} \right)^{1/5} \right]^{-1}
\end{equation}
or can be written in terms of $\Gamma$ as:
\begin{equation}
\boxed{
        \beta_{\text{max}}(\Gamma) \approx \sqrt{\We\mathcal{E}} \left[ 1 + \left( \Lambda_\nu \mathcal{E}^2 (1 + \Gamma) \right)^{1/5} \right]^{-1}.}
    \label{eq:maxspread}
\end{equation}

\subsection{Leading-Order Velocity Closure}

In deriving the scaling, the characteristic velocity within the viscous damping term is evaluated at its leading-order limit, $V \approx \sqrt{\mathcal{E}}$. To justify this mathematically, we substitute the full velocity closure into the viscous term of the timescale interpolation:
\begin{align}
    \left( \Lambda V \mathcal{E}^{3/2} \right)^{1/5} &= \left( \Lambda \left( \sqrt{\mathcal{E}} \big[ T^6 + \Lambda \mathcal{E}^2 T^5 \big]^{-1/6} \right) \mathcal{E}^{3/2} \right)^{1/5} \nonumber \\
    &= (\Lambda \mathcal{E}^2)^{1/5} \big[ T^6 + \Lambda \mathcal{E}^2 T^5 \big]^{-1/30}\\
    &= (\Lambda \mathcal{E}^2)^{1/5}.
\end{align}
Because the dynamic viscous variation is suppressed by the extreme fractional exponent of $-1/30$, the bracketed term has a value between 1 and 1.1 (around 10\% error at its peak) and approaches unity for most of the physical impact parameters. Consequently, the higher-order coupling is mathematically negligible, validating the use of the zeroth-order closure $V \approx \sqrt{\mathcal{E}}$ to achieve an approximate closure. The same argument applies for the elastic term.

\subsection{Asymptotic Convergence to Hertzian Contact Mechanics}

To validate the physical integrity of the dimensionless framework, we test the convergence of the elasto-viscocapillary closure to the classical Hertz theory of elastic contact in the inviscid, zero-tension limit. We begin with the spreading ratio:
\begin{equation}
    \beta_{\text{max}} \approx \sqrt{\We\mathcal{E}} \left[ 1 + \left( \Lambda_{\nu}\mathcal{E}^2 + \Lambda_{el} \mathcal{E}^{3/2} \right)^{1/5} \right]^{-1}
\end{equation}

Expanding the damping parameter ($\Lambda_{eff} \mathcal{E}^2 = \Lambda_{\nu}\mathcal{E}^2 + \Lambda_{el} \mathcal{E}^{3/2}$) and applying the boundary conditions of an ideal Hertzian solid (inviscid: $\mu \to 0, \Lambda_{\nu}\to 0$; zero-tension: $\sigma \to 0, \We\to \infty, \mathcal{E} \to 1$), the expression reduces to:
\begin{equation}
    \beta_{\text{max}} \approx \We^{1/2} \left[ 1 + \left( \We^{5/2} \frac{8}{5\pi} \El \right)^{1/5} \right]^{-1}
\end{equation}

In the extreme elastic damping limit ($\We\to \infty$), the damping term dominates the capillary unity term ($1 \ll \dots$), simplifying the closure to:
\begin{equation}
    \beta_{\text{max}} \approx \frac{\We^{1/2}}{\left( \We^{5/2} \frac{8}{5\pi} \El \right)^{1/5}} = \frac{\We^{1/2}}{\We^{1/2} \left( \frac{8}{5\pi} \El \right)^{1/5}}
\end{equation}
Canceling the inertial scaling $\We^{1/2}$ yields the limit:
\begin{equation}
    \beta_{\text{max}} = \left( \frac{8}{5\pi} \El \right)^{-1/5} \implies 1 = \frac{8}{5\pi} \left( \frac{G}{\rho V_0^2} \right) \beta_{\text{max}}^5
\end{equation}

Multiplying by the kinetic energy ($E_k = \frac{1}{2}\rho \Omega V_0^2$) and substituting the droplet volume ($\Omega = \frac{4}{3}\pi R_0^3$):
\begin{equation}
    \frac{1}{2} m V_0^2 = \frac{4}{5\pi} G \left( \frac{4}{3}\pi R_0^3 \right) \beta_{\text{max}}^5 = \frac{16}{15} G R_0^3 \beta_{\text{max}}^5
\end{equation}

Applying the geometric kinematic constraint $\delta_{\text{max}} = \beta_{\text{max}}^2 R_0$ (which implies $\beta_{\text{max}}^5 = \delta_{\text{max}}^{5/2} / R_0^{5/2}$):
\begin{equation}
    \frac{1}{2} m V_0^2 = \frac{16}{15} G R_0^3 \left( \frac{\delta_{\text{max}}^{5/2}}{R_0^{5/2}} \right) = \frac{16}{15} G R_0^{1/2} \delta_{\text{max}}^{5/2}
\end{equation}

Factoring the Hertzian stiffness $K = \frac{8}{3} G \sqrt{R_0}$ for an incompressible solid ($\nu=0.5$):
\begin{equation}
    \frac{1}{2} m V_0^2 = \frac{2}{5} \left( \frac{8}{3} G \sqrt{R_0} \right) \delta_{\text{max}}^{5/2} = \frac{2}{5} K \delta_{\text{max}}^{5/2}
\end{equation}

This identity reconstructs the energy balance of classical Hertz theory, where initial kinetic energy is equated to the work done against the non-linear restoring force $F = K\delta^{3/2}$. This convergence implies that the closure remains consistent across the transition from fluid-like viscous dissipation to solid-like elastic contact.

\section{Error Analysis}

To quantify the fidelity of the theoretical energy balance, we evaluate the ratio of the experimentally measured maximum spreading ratio ($\beta_{exp}$) to the theoretical prediction ($\beta_{theory}$). 

As shown in Figure \ref{fig:error_residual}, the constant-viscosity Boger fluids and quasi-Boger fluids collapse tightly around the ideal theoretical baseline ($\beta_{exp} / \beta_{theory} = 1$). This confirms that within the margins of standard experimental variance, the elastic stresses do not systematically impede the spreading kinematics. 

\begin{figure}[h]
    \centering
    \pgfplotsset{compat=1.17} 
			
\definecolor{mycolor0}{RGB}{50,155,105}    
\definecolor{mycolor5}{RGB}{213,81,156}    
\definecolor{mycolor10}{RGB}{54,80,163}    
\definecolor{mycolor125}{RGB}{244,196,48}  
\definecolor{mycolor15}{RGB}{245,127,34}   
\definecolor{mycolorSplash}{RGB}{204,204,255} 
\definecolor{mycolorNo}{RGB}{224,224,224}     

\begin{tikzpicture}			
\begin{axis}[			
width=1.4\figW,			
height=1.2\figH,			
at={(0\figW,0\figH)},			
    xlabel={\small $\Lambda_\nu \cdot \mathcal{E}^{2}$},			
    ylabel={\small $\beta_{\text{exp}} / \beta_{\text{theory}}$},			
    xmin=50, xmax=1000000,			
    ymin=0, ymax=2,			
    xmode=log,	
    axis on top,
    legend style={at={(0.00,0.40)}, anchor=north west, draw=none, fill=none, font=\footnotesize},			
    ticklabel style = {font=\footnotesize}, legend cell align=left, legend image post style={scale=0.8}, 			
    row sep=-2pt 
    ]

\fill[mycolor125!50] (20,0.9) rectangle (1000000,1.1);
\addplot[  line width=2.0pt,  domain=0.0001:10E11, samples=200] {1};
\addlegendentry{$\beta_{\text{exp}} / \beta_{\text{theory}} = 1$}

\addplot[only marks, color=teal, mark=triangle*, mark size=3.0pt, line width=0.5pt]			
table[row sep=crcr]{			
12997.1786	1.391015628	\\	
77586.49532	1.355835629	\\	
21960.75006	1.526806512	\\	
131094.4231	1.495940695	\\	
96358.3931	2.060623656	\\	
575210.2239	2.188446314	\\	
6162.455373	1.296187867	\\	
36786.70037	1.27324826	\\	
2577.026792	1.194164808	\\	
15383.52924	1.194786427	\\	
3719.882152	1.257928913	\\	
22205.79004	1.243138781	\\	
8291.303593	1.550913758	\\	
49494.83322	1.547719325	\\	
1490.193754	1.144392299	\\	
8895.692998	1.13127369	\\	
739.4946447	1.064279775	\\	
4414.404044	1.033830814	\\	
934.4523238	1.098424458	\\	
5578.201474	1.079763664	\\	
1546.216075	1.233483499	\\	
9230.117547	1.229048301	\\	
};			
\addlegendentry{Shear-thinning}

\addplot[only marks, color=red, mark=o, mark size=2.5pt, line width=0.5pt]			
table[row sep=crcr]{			
430.2514297	1.001149357	\\	
2568.380535	0.985370522	\\	
504.2008941	1.017215345	\\	
3009.820939	1.007835506	\\	
656.5816088	1.126594988	\\	
3919.455712	1.076016299	\\	
1523.807147	1.139694308	\\	
9096.347727	1.14280313	\\	
318.2067865	0.975165545	\\	
351.8201795	0.976813519	\\	
2100.186166	0.979223199	\\	
851.5392879	1.058023755	\\	
5083.253142	1.054068635	\\	
416.8060725	1.063601906	\\	
2488.118643	1.058547545	\\	
302.5205365	0.959204284	\\	
1805.892564	0.928648824	\\	
307.0023222	0.972031789	\\	
1832.646527	0.929785623	\\	
605.041073	1.028868227	\\	
3611.785127	1.012301025	\\	
272.2684828	0.961322913	\\	
1625.303307	0.949945115	\\	
280.1116079	0.97018941	\\	
1672.122744	0.95112997	\\	
376.9181795	0.956935608	\\	
2250.008364	0.95390193	\\	
302.5205365	1.036936825	\\	
1805.892564	1.038305289	\\	
248.2909292	0.960887927	\\	
1482.1696	0.954862204	\\	
246.2741256	0.979853436	\\	
1470.130317	0.951355375	\\	
302.0723579	0.981998187	\\	
1803.217167	0.957539346	\\	
250.9800006	1.025554699	\\	
1498.221979	1.024695634	\\	
240.2237149	1.013149814	\\	
1434.012465	0.993208913	\\	
244.7055006	0.999052352	\\	
1460.766429	0.99947817	\\	
264.6494471	0.994949785	\\	
1579.821569	0.982301853	\\	
246.4982149	1.026850786	\\	
1471.468015	1.038023843	\\	
246.0500363	1.00758894	\\	
1468.792618	1.011598011	\\	
125.0553845	0.989040709	\\	
583.7163698	0.970889687	\\	
1791.709687	1.013934265	\\	
};			
\addlegendentry{Quasi Boger fluids}

\addplot[only marks, color=blue, mark=square, mark size=2.5pt, line width=0.5pt]			
table[row sep=crcr]{			
18410.88374	0.869213755	\\
88662.26018	0.978731606	\\
273727.0333	1.045167802	\\
5802.421079	1.074678118	\\
27366.08871	0.999805318	\\
84321.53132	0.979838408	\\
2544.787676	1.057237717	\\
11998.51515	0.963566991	\\
36966.33677	0.969868299	\\
1275.592909	0.962349321	\\
6016.904678	0.960134212	\\
18540.43925	0.960528165	\\
125.0553845	0.982636561	\\
583.7163698	0.978939289	\\
1791.709687	1.016863442	\\
};			
\addlegendentry{Boger fluids}

\end{axis}			
\end{tikzpicture}
    \caption{Residual distribution of the maximum spreading ratio. The ratio of experimental to theoretical spreading ($\beta_{\text{exp}} / \beta_{\text{theory}}$) is evaluated against the viscous damping parameter ($\Lambda_\nu \mathcal{E}^2$). Shaded area represents $\pm 10 \%$ deviation.}
    \label{fig:error_residual}
\end{figure}

\clearpage

\bibliographystyle{elsarticle-num} 
\bibliography{bibliography}